\documentstyle[12pt,epsf]{article}
\begin{document}

\title{One-loop counterterms for the dimensional regularization
of arbitrary Lagrangians.}

\author{Petr Pronin
\thanks{E-mail:$pronin@theor.phys.msu.su$}
and Konstantin Stepanyantz
\thanks{E-mail:$stepan@theor.phys.msu.su$}\\
{\small\em Moscow State University, Physical Faculty,}\\
{\small\em Department of Theoretical Physics.}\\
{\small\em $117234$, Moscow, Russian Federation}}

\maketitle

\begin{abstract}
We present master formulas for the divergent part of the one-loop
effective action for an arbitrary (both minimal and nonminimal)
operators of any order in the 4-dimensional curved space. They can be
considered as computer algorithms, because the one-loop calculations
are then reduced to the simplest algebraic operations. Some test
applications are considered by REDUCE analitical calculation system.
\end{abstract}

\begin{center}
pacs nombers 11.10.Gh, 04.62.+v
\end{center}

\unitlength=1pt
\sloppy


\section{Introduction.}
\hspace{\parindent}

Progress in the quantum field theory and quantum gravity in particular
depends much on the development of methods for the calculation of the
effective action. For a lot of problems the analyses can be confined to the
one-loop approximation. In this case

\begin{eqnarray}\label{efac}
\Gamma[\varphi] =  S[\varphi]
+ \frac{i}{2}\ \hbar\ \mbox{tr}\ \ln\ D + O(\hbar^2),\quad \mbox{where}\quad
D_i{}^j \equiv \frac{\delta^2 S}{\delta \varphi^i\ \delta \varphi_j}.
\end{eqnarray}

\noindent
where latin letters denote the whole set of $\varphi$ indexes.

Finding its divergent part is sometimes a rather complicated technical
problem, especially in the curved space. Unfortunately, the usual diagram
technique is not manifestly covariant. A very good tool to make
calculations in the curved space-time is the Schwinger-DeWitt proper time
method \cite{schwinger,dewitt}. It allows to make manifestly covariant
calculations of Feynman graphs if the propagator depends on the background
metric \cite{barv}. This approach was successfully applied to obtain
one-loop counterterms in theories with the simplest second and forth
order operators $D_i{}^j$. We should also mention other covariant methods,
that in principle allow to get the divergent part of effective action
\cite{lps,gusynin,gusynin1}.

A different approach was proposed by t'Hooft and Veltman \cite{thooft}.
Instead of calculating Feynman graphs for each new theory, they made it
only once for a rather general case. Nevertheless, if $D_i{}^j$ in
(\ref{efac}) is not a minimal second order oparetor, their results can
not be used directly.

In this paper we extend t'Hooft and Veltman approach to the most
general case. We construct the explicit expression for the divergent part
of the one-loop effective action without any restrictions to the form and
the order of the operator $D_i{}^j$ in the 4-dimentional space-time.
Then the divergent part of the one-loop effective action can be found
only by making the simplest algebraic operations, for example, by computers.

Our paper is organized as follows.

In Sec. 2 we briefly remind t'Hooft-Veltman diagram technique and
introduce some notations and definitions.

In Sec. 3 we consider a theory with an arbitrary minimal operator
in the curved space-time. The main result here is an explicit expression
for the one-loop contribution to the divergent part of the effective action.

In Sec. 4  we describe a method for the derivation of a master formula
for an arbitrary nonminimal operator on the curved background and present
the result.

Sec. 5 is devoted to the consideration of some particular cases. We show
the agreement of our results with the earlier known ones. Here we proove
the correctness of the method and that is why we do not consider here new
applications.

In Sec. 6 we give a summary of our results and discuss prospects
of using computers for the automatization of calculations.

In Appendix A we describe in details the derivation of the one-loop
counterterms for an arbitrary minimal operator.

In Appendix B we illustrate the general method presented in Sec. 4 by
calculating the simplest Feynman graphs for an arbitrary nonminimal
operator.

Appendix C is devoted to the derivation of some useful identities.

\section{Diagramic approach in the background field method}
\hspace{\parindent}

We will calculate the divergent part of the one-loop effective action
for a general theory by the diagram tecknique. First we note, that

\begin{eqnarray}
D_i{}^j = \frac{\delta^2 S}{\delta \varphi^{i}\delta \varphi_j}
\end{eqnarray}

\noindent
is a differential operator depending on the background field $\varphi$.
Its most general form is

\begin{eqnarray}\label{nonminimal}
&&
\vphantom{\frac{1}{2}}
D_{i}{}^{j} =
K^{\mu_1\mu_2\ldots \mu_{L}}{}_{i}{}^{j}
\ \nabla_{\mu_1} \nabla_{\mu_2}\ldots
\nabla_{\mu_{L}}
+\ S^{\mu_1\mu_2\ldots \mu_{L-1}}{}_{i}{}^{j}
\ \nabla_{\mu_1} \nabla_{\mu_2}\ldots
\nabla_{\mu_{L-1}}
\nonumber\\
&&
\vphantom{\frac{1}{2}}
+\ W^{\mu_1\mu_2\ldots \mu_{L-2}}{}_{i}{}^{j}
\ \nabla_{\mu_1} \nabla_{\mu_2}\ldots
\nabla_{\mu_{L-2}}
+\ N^{\mu_1\mu_2\ldots \mu_{L-3}}{}_{i}{}^{j}
\ \nabla_{\mu_1} \nabla_{\mu_2}\ldots
\nabla_{\mu_{L-3}}
\nonumber\\
&&
\vphantom{\frac{1}{2}}
+\ M^{\mu_1\mu_2\ldots \mu_{L-4}}{}_{i}{}^{j}
\ \nabla_{\mu_1} \nabla_{\mu_2}\ldots
\nabla_{\mu_{L-4}} +
\ldots,
\end{eqnarray}

\noindent
where $\nabla_\mu$ is a covariant derivative:

\begin{eqnarray}
&&\nabla_\alpha T^\beta{}_i{}^j = \partial_\alpha T^\beta{}_i{}^j +
\Gamma_{\alpha\gamma}^\beta T^\gamma{}_i{}^j + \omega_{\alpha i}{}^k
T^\beta{}_k{}^j - T^\beta{}_i{}^k \omega_{\alpha k}{}^j;\nonumber\\
&&\vphantom{\frac{1}{2}}
\nabla_\mu \Phi_i = \partial_\mu \Phi_i + \omega_{\mu i}{}^j \Phi_j.
\end{eqnarray}

Here $\Gamma_{\mu\nu}^\alpha$ is the Cristoffel symbol

\begin{equation}
\Gamma_{\mu\nu}^\alpha = \frac{1}{2} g^{\alpha\beta} (\partial_\mu
g_{\nu\beta} + \partial_\nu g_{\mu\beta} - \partial_\beta g_{\mu\nu})
\end{equation}

\noindent
and $\omega_{\mu i}{}^j$ is a connection on the principle bundle.

Commuting covariant derivatives we can always make $K, S, W, N, M, \ldots$
symmetric in the greek indexes. This condition is very convenient for
the calculations, so we will assume it to be satisfied.

The operator is called minimal if $L = 2l$ and
$K^{\mu\nu\ldots\alpha}{}_i{}^j =
K_0{}^{\mu\nu\ldots\alpha}\delta_i{}^j,$
where $K_0$ is a totally symmetric tensor, built by $g_{\mu\nu}$:

\begin{equation}\label{k0}
K_0{}^{\mu\nu\alpha\beta \dots } = \frac{1}{(2l-1)!!}
(g_{\mu\nu} g_{\alpha\beta} \ldots  + \mbox{permutations of}
\quad (\mu\nu\alpha\beta\ldots) ).
\end{equation}

\noindent
If an operator can not be reduced to this form, we will call it
nonminimal one.

Commuting covariant derivatives we can rewrite a minimal operator in
the form:

\begin{eqnarray}\label{minimal1}
&&D_{i}{}^{j}\ =\ \delta_{i}{}^{j}\ \Box^l +
\ S^{\mu_1\mu_2\ldots \mu_{2l-1}}{}_{i}{}^{j}
\ \nabla_{\mu_1} \nabla_{\mu_2}\ldots
\nabla_{\mu_{2l-1}}
+\ W^{\mu_1\mu_2\ldots \mu_{2l-2}}{}_{i}{}^{j}
\nonumber\\
&&\vphantom{\frac{1}{2}}
\times
\nabla_{\mu_1} \nabla_{\mu_2}\ldots
\nabla_{\mu_{2l-2}}
+\ N^{\mu_1\mu_2\ldots \mu_{2l-3}}{}_{i}{}^{j}
\nabla_{\mu_1} \nabla_{\mu_2}\ldots
\nabla_{\mu_{2l-3}}
\nonumber\\
&&\vphantom{\frac{1}{2}}
+\ M^{\mu_1\mu_2\ldots \mu_{2l-4}}{}_{i}{}^{j}
\nabla_{\mu_1} \nabla_{\mu_2}\ldots
\nabla_{\mu_{2l-4}} +
\ldots,
\end{eqnarray}

\noindent
where $\Box \equiv \nabla_{\mu} \nabla^{\mu}$.


The one-loop effective action (\ref{efac}) can be presented as a sum of
one-loop diagrams, say, for a minimal operator, as follows:

\begin{eqnarray}\label{purttheor}
&&\Gamma^{(1)} = \frac{i}{2} \mbox{tr}\ \ln\ D_i{}^j =
\frac{i}{2} \mbox{tr}\ \ln\ \left(\partial^{2l} + V \right)
=\frac{i}{2} \mbox{tr}\ \ln\ \partial^{2l}
\nonumber\\
&&
+\ \frac{i}{2} \mbox{tr}\ \ln\ \left(1 + \frac{1}{\partial^{2l}} V\right)
= \frac{i}{2} \mbox{tr} \sum_{k=1}^\infty \frac{1}{k}
\left(-\ \frac{1}{\partial^{2l}} V\right)^k.
\end{eqnarray}

\noindent
where

\begin{eqnarray}\label{purt}
&&\vphantom{\frac{1}{2}}
V = S^{\mu_1\ldots\mu_{L-1}} \partial_{\mu_1}\ldots\partial_{\mu_{L-1}}
+\ W^{\mu_1\ldots\mu_{L-2}} \partial_{\mu_1}\ldots\partial_{\mu_{L-2}}
+\ N^{\mu_1\ldots\mu_{L-3}}
\nonumber\\
&&\vphantom{\frac{1}{2}}
\times \partial_{\mu_1}\ldots\partial_{\mu_{L-3}}
+\ M^{\mu_1\ldots\mu_{L-4}} \partial_{\mu_1}\ldots\partial_{\mu_{L-4}}
+ \ldots + O(h,\omega).
\end{eqnarray}

\noindent
and we omit an infinite numerical constant.

Terms $O(h,\omega)$ can be found by a series expansion of the operator
in powers of $h_{\mu\nu} = g_{\mu\nu} - \eta_{\mu\nu}$ and
$\omega_\mu{}_i{}^j$.

In the momentum space the propagator is $\delta_{i}^{j} /k^{2l}$.
The form of the vertexes is rather evident, for example, the vertex with
the external line of $S$ type can be written as

\begin{eqnarray}
S^{\mu_1\mu_2\ldots\mu_{L-1}} k_{\mu_1} k_{\mu_2} \ldots k_{\mu_{L-1}}
\equiv (Sk).
\end{eqnarray}

Similar notations we will use for other expressions, for example,

\begin{eqnarray}
&&(W(k+p))^\alpha \equiv
W^{\mu\nu\ldots \beta\alpha} (k+p)_\mu (k+p)_\nu\ldots (k+p)_\beta,
\nonumber\\
&&(Sk) ^{\alpha\beta}\ \equiv
S^{\mu\nu\ldots \gamma\alpha\beta} k_\mu k_\nu\ldots k_\gamma.
\end{eqnarray}

Numerical factors for the Feynman graphs can be easily found by
(\ref{purttheor}).

The number of diagrams in (\ref{purttheor}) is infinite, but
most of them are convergent. Really, it is easy to see that
the degree of divergence of a one-loop graph with $s$ legs of $S$ type,
$w$ legs of $W$ type, $n$ legs of $N$ type, $m$ legs of $M$ type and
so on ($k = s+w+n+m+\ldots$) in the flat space ($h_{\mu\nu}=0$,
$\omega_\mu{}_i{}^j=0$) is

\begin{eqnarray}\label{degree}
I = 4 - s - 2 w - 3 n -4 m - \ldots.
\end{eqnarray}

\noindent
Therefore, there are only a finite number of the divergent diagrams.
They are presented at the Fig.~\ref{flatdiagrams}. (We excluded divergent
graphs, that give zero contribution to the effective action, for example,
some tadpole ones.)

The extension of this results to a nonminimal operator will be made below.


\section{Effective action for the theory with minimal operator.}
\hspace{\parindent}

\begin{figure}[l]
\hspace*{-3cm}
\epsfbox{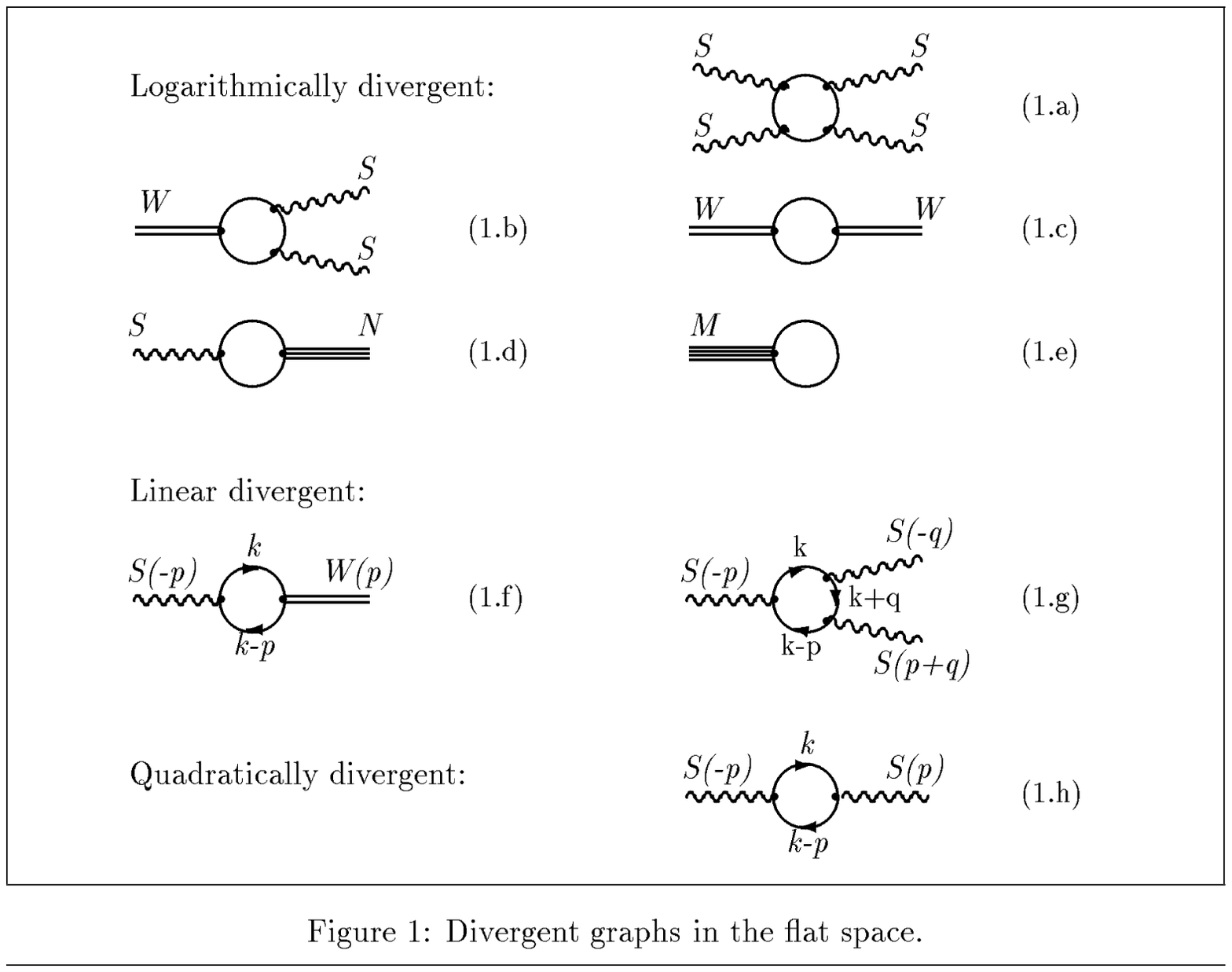}
\label{flatdiagrams}
\end{figure}

Now we should calculate the divergent part of diagrams presented at the
Fig.~\ref{flatdiagrams}. We will do it using dimensional regularization.
So, in order to find the divergent part of an integral

\begin{equation}
\int d^{d}k\ f(k,p)
\end{equation}

\noindent
it is necessary to expand the function $f$ into series, retain only
logarithmically divergent terms and perform the integration according
to the following equations \cite{hep1}

\begin{eqnarray}\label{angleintegration}
&&\int d^{d}k\ \frac{1}{k^{2m+5}}
k_{\mu_{1}} k_{\mu_{2}} \ldots  k_{\mu_{2m+1}} = 0,\nonumber\\
&&\int d^{d}k\ \frac{1}{k^{2m+4}}
k_{\mu_{1}} k_{\mu_{2}} \ldots  k_{\mu_{2m}}
= -\ \frac{2 i\pi^2}{(d-4)} <n_{\mu_1} n_{\mu_2} \ldots n_{\mu_{2m}}>,
\end{eqnarray}

\noindent
where $n_\mu$ is a unit vector ($n_\mu n^\mu = 1$) and

\begin{eqnarray}\label{angle}
&&<n_{\mu_{1}} n_{\mu_{2}} \ldots n_{\mu_{2m}}>\ \equiv
\frac{1}{2^{m} (m+1)!}
\nonumber\\
\vphantom{\frac{1}{2}}
&&\times
\left(g_{\mu_{1}\mu_{2}} g_{\mu_{3}\mu_{4}} \ldots g_{\mu_{2m-1}\mu_{2m}}
+ \mbox{permutaions of}\ (\mu_1 \ldots \mu_{2m})
\right)
\end{eqnarray}

Using this prescription one can easily find the divergent part of the
diagrams at the Fig~\ref{flatdiagrams}. The calculations are presented
in details in the appendix \ref{minimalcalc}.

Collecting the results for all graphs, we obtain the divergent part of
the one-loop effective action for the minimal operator (\ref{minimal1})
in the flat space:

\begin{eqnarray}\label{flat}
&&\left( \Gamma^{(1)}_\infty \right)^{flat}=
\frac{1}{16\pi^2(d-4)}\ \mbox{tr} \int d^4 x
<(l-1) \partial_\mu \hat S\ \hat W^\mu
- \frac{(2l-1)}{2} \partial_\mu \hat S^\mu \hat W
\nonumber\\
&&
+\frac{l^2}{2(2l+1)} \partial_\mu \hat S\ \partial^\mu \hat S
-\frac{(2l-1)(l^2-1)}{2(2l+1)}
\partial_\mu \hat S^{\mu\nu} \partial_\nu \hat S
+\frac{(2l-1)^2 l}{4(2l+1)} \partial_\mu \hat S^\mu \partial_\nu \hat S^\nu
\nonumber\\
&&
\vphantom{\frac{1}{2}}
+\frac{(2l-1)}{3} \partial_\mu \hat S^\mu \hat S\ \hat S
-\frac{(2l-1)}{3} \partial_\mu \hat S\ \hat S\ \hat S^\mu
+\hat S\ \hat N + \frac{1}{2} \hat W^2 + \frac{1}{4} \hat S^4
- \hat W\ \hat S^2
\nonumber\\
&&\vphantom{\frac{1}{2}}
-\hat M>.
\end{eqnarray}

\noindent
where

\begin{eqnarray}\label{minnot}
&&\hat N = (Nn) = N^{\mu\nu\ldots \alpha} n_\mu n_\nu\ldots n_\alpha;
\nonumber\\
&&\vphantom{\frac{1}{2}}
\hat S^\mu = (Sn)^\mu = S^{\mu\nu\ldots \alpha} n_\nu\ldots n_\alpha
\quad \mbox{and so on}.
\end{eqnarray}

In order to extend this result to the curved space-time we first
consider a minimal operator in the form (\ref{minimal1}).

In this case we can not calculate all divergent graphs, because
their number is infinite. (The matter is that the degree of divergence
does not depend on the number of $h_{\mu\nu}$ vertexes and there are
infinite number of such vertexes too). Nevertheless if we note that the
answer should be invariant under the general coordinate transformations,
the result can be found by calculating only a finite number of graphs.
Really, we should replace derivatives in (\ref{flat}) by the covariant
ones and add expressions, containing curvature tensors
$R^\sigma{}_{\alpha\mu\nu}$ and $F_{\mu\nu}$. The most general form of
additional terms is

\begin{eqnarray}\label{anzats}
&&\vphantom{\frac{1}{2}}
\frac{1}{16\pi^2(d-4)}\ \mbox{tr} \int d^4 x\ \sqrt{-g}
< a_1 R^2 + a_2 R_{\mu\nu} R^{\mu\nu}
+ a_3 \hat W^{\alpha\beta} R_{\alpha\beta}
+ a_4 \hat W R
\nonumber\\
&&\vphantom{\frac{1}{2}}
+ a_5 \nabla_\mu \hat S^{\mu\alpha\beta} R_{\alpha\beta}
+ a_6 \nabla_\mu \hat S^{\mu} R
+ a_7 \hat S^2 R
+ a_8 R_{\alpha\beta} \hat S^\alpha \hat S^\beta
+ a_9 R_{\alpha\beta} \hat S^{\alpha\beta} \hat S
\nonumber\\
&&\vphantom{\frac{1}{2}}
+ a_{10} R_{\mu\nu\alpha\beta} \hat S^{\mu\alpha} \hat S^{\nu\beta}
+ a_{11} F_{\mu\nu} F^{\mu\nu}
+ a_{12} F_{\mu\nu} \hat S^\mu \hat S^\nu
+ a_{13} F_{\mu\nu} \nabla^\mu \hat S^\nu
>.
\end{eqnarray}

\noindent
where

\begin{eqnarray}
&&\vphantom{\frac{1}{2}}
R^{\alpha}{}_{\beta\mu\nu} =
\partial_\mu \Gamma_{\nu\beta}^{\alpha} -
\partial_\nu \Gamma_{\mu\beta}^{\alpha} +
\Gamma_{\mu\gamma}^{\alpha} \Gamma_{\nu\beta}^{\gamma} -
\Gamma_{\nu\gamma}^{\alpha} \Gamma_{\mu\beta}^{\gamma},\nonumber\\
&&\vphantom{\frac{1}{2}}
F_{\mu\nu i}{}^j =
\partial_\mu \omega_{\nu i}{}^j -
\partial_\nu \omega_{\mu i}{}^j +
\omega_{\mu i}{}^k \omega_{\nu k}{}^j -
\omega_{\nu i}{}^k \omega_{\mu k}{}^j,\nonumber\\
&&\vphantom{\frac{1}{2}}
R_{\mu\nu} = R^{\alpha}{}_{\mu\alpha\nu},
\qquad R = g^{\mu\nu} R_{\mu\nu}.
\end{eqnarray}

\noindent
(We take into account that the expression
$R_{\mu\nu\alpha\beta} R^{\mu\nu\alpha\beta} -
4 R_{\mu\nu} R^{\mu\nu} + R^2$
is a total derivative and may be omitted).

Then the coefficients $a_1$ - $a_{13}$ can be found by calculating
the diagrams presented at the Fig.~\ref{curveddiagrams}. They conform
to the first nontrivial approximation in the counterterm expansion
in powers of weak fields $h_{\mu\nu}=g_{\mu\nu}-\eta_{\mu\nu}$ and
$\omega_{\mu i}{}^j$.

In the appendix \ref{minimalcalc} we present detailed calculation of the
coefficients $a_3$ and $a_4$. The other ones were found in the same way.
After rather cumbersome calculations we obtain the following formula for
the divergent part of the one-loop effective action for a minimal operator
(\ref{minimal1}) in the curved space:

\begin{eqnarray}\label{diaganswer1}
&&
\Gamma^{(1)}_\infty=
\frac{1}{16\pi^2(d-4)}\ \mbox{tr} \int d^4 x\ \sqrt{-g}
<(l-1)\ \nabla_\mu \hat S\ \hat W^\mu
- \frac{(2l-1)}{2} \nabla_\mu \hat S^\mu \hat W
\nonumber\\
&&
-\frac{(2l-1)(l^2-1)}{2(2l+1)}
\nabla_\mu \hat S^{\mu\nu} \nabla_\nu \hat S
+\frac{l^2}{2(2l+1)} \nabla_\mu \hat S\ \nabla^\mu \hat S
+\frac{(2l-1)^2 l}{4(2l+1)} \nabla_\mu \hat S^\mu
\nonumber\\
&&
\times \nabla_\nu \hat S^\nu
+\frac{(2l-1)}{3} \nabla_\mu \hat S^\mu \hat S\ \hat S
-\frac{(2l-1)}{3} \nabla_\mu \hat S\ \hat S\ \hat S^\mu
+\hat S\ \hat N + \frac{1}{2} \hat W^2 + \frac{1}{4} \hat S^4
\nonumber\\
&&
- \hat W\ \hat S^2-\hat M
- \frac{(2l-3)(l-1)}{6}\ \hat W^{\alpha\beta} R_{\alpha\beta}
+ \frac{l}{6} \hat W R
- \frac{(2l-1)l}{12}\ \nabla_\mu \hat S^\mu R
\nonumber\\
&&
+ \frac{(2l-1)(2l-3)(l-1)}{12}\ \nabla_\mu \hat S^{\mu\alpha\beta}
R_{\alpha\beta}
- \frac{l}{6} \hat S^2 R
+ \frac{(2l-1)^2 (l+2)}{24(2l+1)}\ R_{\alpha\beta}
\nonumber\\
&&
\hat S^\alpha \hat S^\beta + \frac{(2l-1)(l-1)}{6} R_{\alpha\beta}
\hat S^{\alpha\beta} \hat S
-\frac{(2l-1)(l-1)^2 (l+2)}{12(2l+1)} R_{\mu\nu\alpha\beta}
\hat S^{\mu\alpha} \hat S^{\nu\beta}
\nonumber\\
&&
+ l (\frac{1}{120} R^2 + \frac{1}{60} R_{\mu\nu} R^{\mu\nu})
+ \frac{l(2l-1)}{6}\nabla_\mu \hat S_\nu F^{\mu\nu}
+ \frac{(2l-1)^2 (l+1)}{4(2l+1)} F_{\mu\nu} \hat S^\mu \hat S^\nu
\nonumber\\
&&
+ \frac{l}{12} F_{\mu\nu} F^{\mu\nu}>.
\end{eqnarray}

It is more convenient sometimes to use a different form of the operator:

\begin{eqnarray}\label{minimal2}
&&
\vphantom{\frac{1}{2}}
D_{i}{}^{j} = \delta_{i}{}^{j}
K_0 {}^{\mu_1\mu_2\ldots \mu_{2l}}
\nabla_{\mu_1} \nabla_{\mu_2}\ldots
\nabla_{\mu_{2l}}
+\ S^{\mu_1\mu_2\ldots \mu_{2l-1}}{}_{i}{}^{j}
\ \nabla_{\mu_1} \nabla_{\mu_2}\ldots
\nabla_{\mu_{2l-1}}
\nonumber\\
&&
\vphantom{\frac{1}{2}}
+\ W^{\mu_1\mu_2\ldots \mu_{2l-2}}{}_{i}{}^{j}
\ \nabla_{\mu_1} \nabla_{\mu_2}\ldots
\nabla_{\mu_{2l-2}}
+\ N^{\mu_1\mu_2\ldots \mu_{2l-3}}{}_{i}{}^{j}
\ \nabla_{\mu_1} \nabla_{\mu_2}\ldots
\nabla_{\mu_{2l-3}} \nonumber\\
&&+
\vphantom{\frac{1}{2}}
\ M^{\mu_1\mu_2\ldots \mu_{2l-4}}{}_{i}{}^{j}
\ \nabla_{\mu_1} \nabla_{\mu_2}\ldots
\nabla_{\mu_{2l-4}} +
\ldots,
\end{eqnarray}

\noindent
where $K_0$ was defined in (\ref{k0}).

By the help of the same method we found the following answer for
the divergent part of the one-loop effective action:

\begin{eqnarray}\label{diaganswer2}
&&
\Gamma^{(1)}_\infty=
\frac{1}{16\pi^2(d-4)}\ \mbox{tr} \int d^4 x\ \sqrt{-g}
<(l-1)\ \nabla_\mu \hat S\ \hat W^\mu
- \frac{(2l-1)}{2} \nabla_\mu \hat S^\mu \hat W
\nonumber\\
&&
+\frac{l^2}{2(2l+1)} \nabla_\mu \hat S\ \nabla^\mu \hat S
+\frac{(2l-1)^2 l}{4(2l+1)} \nabla_\mu \hat S^\mu \nabla_\nu \hat S^\nu
-\frac{(2l-1)(l^2-1)}{2(2l+1)}
\nabla_\mu \hat S^{\mu\nu}
\nonumber\\
&&
\times \nabla_\nu \hat S
+\frac{(2l-1)}{3} \nabla_\mu \hat S^\mu \hat S\ \hat S
-\frac{(2l-1)}{3} \nabla_\mu \hat S\ \hat S\ \hat S^\mu
+\hat S\ \hat N + \frac{1}{2} \hat W^2 + \frac{1}{4} \hat S^4
\nonumber\\
&&
- \hat W\ \hat S^2-\hat M
- \frac{l^2(2l-1)}{6(l+1)} \nabla_\mu \hat S^\mu R
+ \frac{(2l-1)(2l-3)(l-1)}{6(l+1)} \nabla_\mu
\hat S^{\mu\alpha\beta} R_{\alpha\beta}
\nonumber\\
&&
+ \frac{l^2 (2l-1)}{3(l+1)} F^{\mu\nu} \nabla_\mu \hat S_\nu
+ \frac{l^2}{3(l+1)} \hat W\ R
- \frac{(2l-3)(l-1)}{3(l+1)} \hat W^{\mu\nu} R_{\mu\nu}
\nonumber\\
&&
+ \frac{(2l-1)(l-1)(l+2)}{6(2l+1)} R_{\mu\nu}
\hat S^{\mu\nu} \hat S
- \frac{(2l-1)(l-1)^2 (l+2)}{12(2l+1)} R_{\mu\nu\alpha\beta}
\hat S^{\mu\alpha} \hat S^{\nu\beta}
\nonumber\\
&&
- \frac{(2l-1)^2 (l-4)}{24(2l+1)} R_{\mu\nu} \hat S^\mu \hat S^\nu
- \frac{l^2}{2(2l+1)} \hat S^2 R
+\frac{(l+1)(2l-1)^2}{4(2l+1)} \hat S^\mu \hat S^\nu F_{\mu\nu}
\nonumber\\
&&
+ \frac{l^2(-7l^2+4l+12)}{540} R_{\mu\nu} R^{\mu\nu}
+ \frac{l^2(3l^2+4l+2)}{1080} R^2
+ \frac{l^3}{12} F^{\mu\nu} F_{\mu\nu}
>.
\end{eqnarray}


\section{Effective action for theory with nonminimal operator.}
\hspace{\parindent}\label{nonmin}

Let us consider a theory with an arbitrary nonminimal operator
(\ref{nonminimal}) first in the flat space ($g_{\mu\nu}= \eta_{\mu\nu}$,
$\omega_\mu{}_i{}^j=0$). In this case

\begin{eqnarray}
\Gamma^{(1)} = \frac{i}{2}\ \mbox{tr}\ \ln\ D_i{}^j =
\frac{i}{2}\ \mbox{tr}\ \ln\ \left((K\partial) + V \right)
= \frac{i}{2}\ \mbox{tr} \sum_{k=1}^\infty \frac{1}{k}
\left(-\ \frac{1}{\partial^{2l}} V\right)^k,
\end{eqnarray}

\noindent
where $V$ is given in (\ref{purt}). There are the following differences
from the minimal operator:

1. For the nonminimal operator the propagator in the momentum space is
$(Kk)^{-1}{}_i{}^j$, where

\begin{eqnarray}
&&(Kk)_i{}^j\ \equiv K^{\mu\nu\ldots \alpha}{}_i{}^j
\ k_\mu k_\nu\ldots k_\alpha;\nonumber\\
&&
\vphantom{\frac{1}{2}}
(Kk)^{-1}{}_i{}^m\ (Kk)_m{}^j = \delta_i{}^j.
\end{eqnarray}

\noindent
(We assume, that $(Kk)^{-1}{}_{i}{}^{j}$ exists. Usually it can be
made by adding gauge fixing terms to the action).

2. For a nonminimal operator $V$ depends on some additional fields $\phi^i$
besides $h_{\mu\nu}$ and $\omega_\mu{}_i{}^j$. It will be considered in
details below.

The divergent graphs are the same as in the case of a minimal operator
(see Fig.~\ref{flatdiagrams}), but because of the difference of the
propagators the calculations are also different. They are considered
in the appendix \ref{nondiagcalc}. The result is

\begin{eqnarray}\label{nondiagflat}
&&
\left( \Gamma^{(1)}_\infty \right)^{flat} = \frac{1}{16\pi^2(d-4)}
\mbox{tr} \int d^4 x <
\frac{1}{4} \hat S^4
- \hat W\ \hat S^2
+ \frac{1}{2} \hat W^2
+ \hat S\ \hat N
- \hat M
\nonumber\\
&&
+\frac{1}{3} \left(
\vphantom{\frac{1}{2}}
(L-1) \partial_\mu \hat S^\mu \hat S^2
- L \partial_\mu \hat S\ \hat K^\mu \hat S^2
- (L-1) \partial_\mu \hat S\ \hat S\ \hat S^\mu
+ L \partial_\mu \hat S\ \hat S^2 \hat K ^\mu
\right)
\nonumber\\
&&
- \frac{1}{2} \partial_\mu \hat S
\ \partial_\nu \hat S
\left( - \frac{1}{2} L(L-1) \hat K ^{\mu\nu}
+ L^2 \hat K^\mu \hat K^\nu \right)
+ \frac{1}{2}L(L-1) \partial_\mu \hat S\ \partial_\nu
\hat S ^\nu \hat K^\mu
\nonumber\\
&&
-\frac{1}{4}(L-1)(L-2) \partial_\mu \hat S\ \partial_\nu
\hat S ^{\mu\nu}
- L \partial_\mu \hat S\ \hat W \hat K ^\mu
+(L-2) \partial_\mu \hat S\ \hat W ^\mu>.
\end{eqnarray}

\noindent
where we use the following notations (compare with (\ref{minnot}))

\begin{eqnarray}\label{ndnotations}
\begin{array}{ll}
\hat W\ \equiv (Kn)^{-1} (Wn);
&\hat K^\alpha\ \equiv (Kn)^{-1} (Kn)^\alpha;\\
(Kn)\ \equiv K^{\mu\nu\ldots \beta} n_\mu n_\nu\ldots n_\beta;
\qquad
&(Kn)^{-1}{}_i{}^m\ (Kn)_m{}^j = \delta_i{}^j
\end{array}
\end{eqnarray}
etc.

The generalization of this result to the curved space-time can also be
made in the frames of the weak field approximation. We should substitute
derivatives in (\ref{nondiagflat}) by the covariant ones and add some
terms, containing curvature tensors. The additional terms can be found
by calculating Feynman diagrams, that conform to the first terms of
their expansion over weak fields. Nevertheless, in this case there
are some difficulties, for example, now we do not know
$K^{\mu\nu\ldots\alpha}{}_i{}^j$ and, therefore, how it depends on
$h_{\mu\nu}$. Thus, we do not know expressions for vertexes in the
weak field limit.

In order to overcome this difficulty we will use the following trick.
Suppose, that $K^{\mu\nu\ldots\sigma}{}_i{}^j$ does not depend on
$g_{\mu\nu}$, but depend on some external fields $\phi^b$. Also,
we impose a condition

\begin{eqnarray}\label{covder}
&&
0 = \nabla_\alpha K^{\mu\nu\ldots\sigma}{}_i{}^j =
\frac{\partial K^{\mu\nu\ldots\sigma}{}_i{}^j}{\partial \phi^b}
\partial_\alpha \phi^b +
\Gamma_{\alpha\beta}^\mu K^{\beta\nu\ldots\sigma}{}_i{}^j + \ldots
\nonumber\\
&&\vphantom{\frac{1}{2}}
+\Gamma_{\alpha\beta}^\sigma K^{\mu\nu\ldots\beta}{}_i{}^j
+ \omega_{\alpha i}{}^k
K^{\mu\nu\ldots\sigma}{}_k{}^j -
K^{\mu\nu\ldots\sigma}{}_i{}^k \omega_{\alpha k}{}^j,
\end{eqnarray}

\noindent
that is, of cause, satisfied if $K^{\mu\nu\ldots\sigma}{}_i{}^j$
depends only on $g_{\mu\nu}$. From (\ref{covder}) we conclude
that $\omega_{\alpha i}{}^j$ should be considered as a weak field.

\begin{figure}[l]
\hspace*{-3cm}
\epsfbox{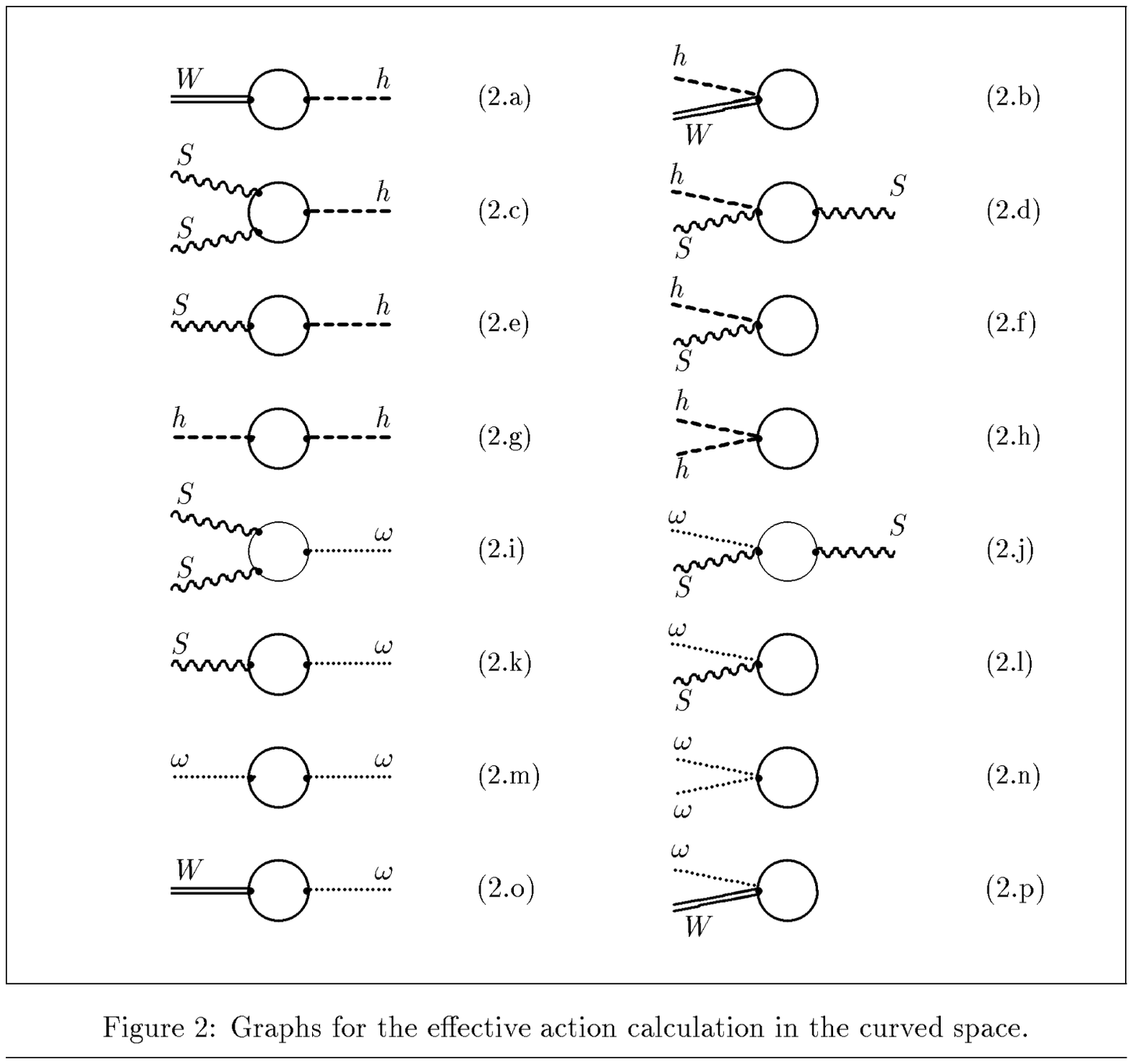}
\label{curveddiagrams}
\end{figure}

\begin{figure}[l]
\hspace*{-3cm}
\epsfbox{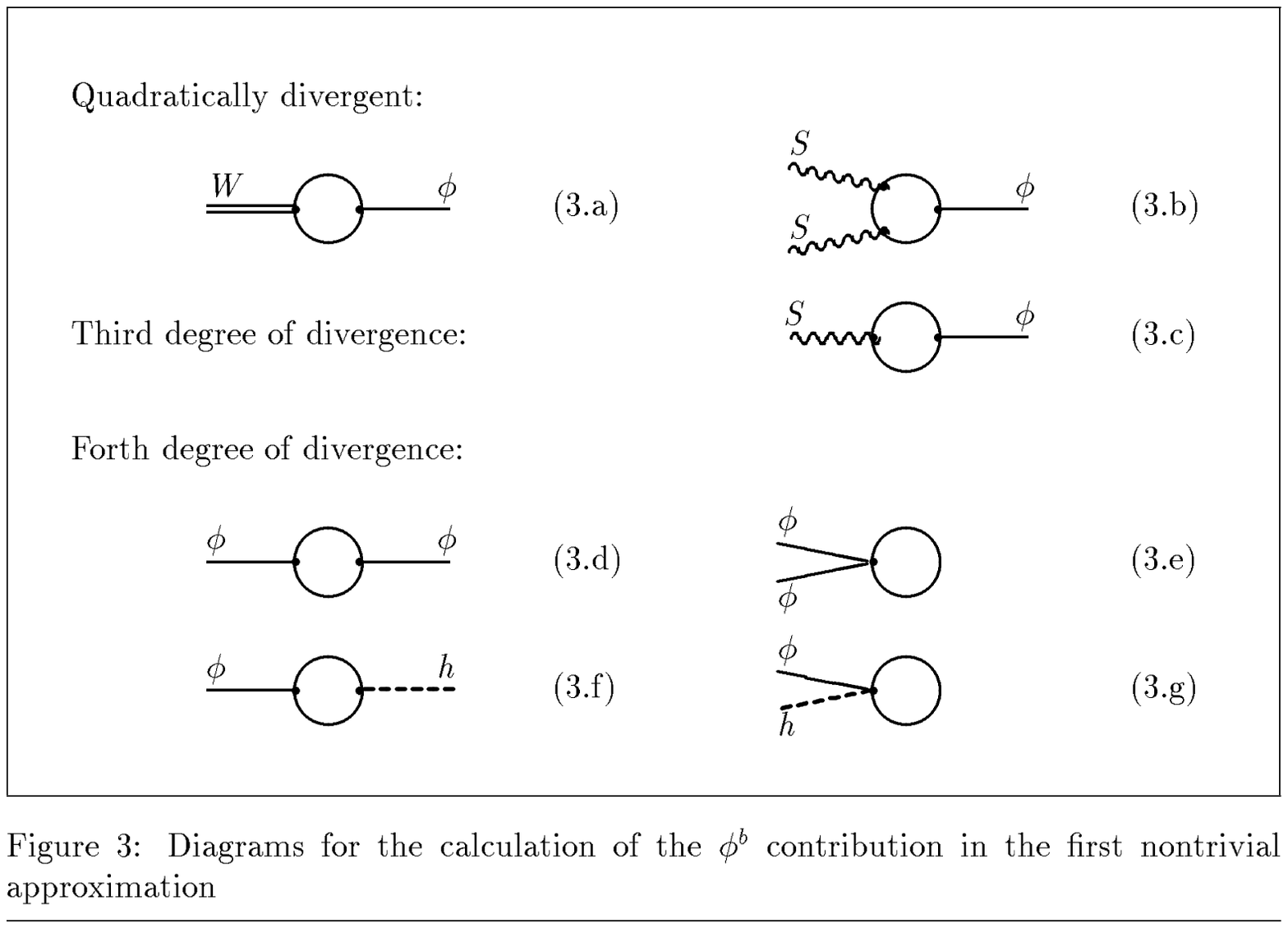}
\label{phidiagrams}
\end{figure}

Thus, unlike the minimal operator, besides the diagrams with external
$h$ and $\omega$ lines, presented at the Fig~\ref{curveddiagrams} we
should consider also graphs with external $\phi$ lines, presented at the
Fig~\ref{phidiagrams}.

Computing this diagrams we obtain expressions, containing
$\partial_\alpha K^{\mu\nu\ldots\sigma}{}_i{}^j$. Substituting it by
(\ref{covder}), we found the result depending on $h_{\mu\nu}$ and
$\omega_\mu{}_i{}^j$. The calculation of graphs, presented at the
Fig~\ref{curveddiagrams} gives results, that can not be written as
a weak field limit of a covariant expression. Nevertheless, the
covariant result should be found by adding a contribution of diagrams
with external $\phi$-lines.

Therefore, for a nonminimal operator we can not calculate terms,
containing $R^\alpha{}_{\beta\mu\nu}$ and $F_{\mu\nu}$ separately.

We illustrate the above discussion in the appendix \ref{nondiagcalc}
by calculating the simplest group of diagrams.

Summing up the results for all graphs, we find the divergent part of
the one-loop effective action for an arbitrary nonminimal operator

\begin{eqnarray}\label{answer}
\Gamma^{(1)}_\infty=
\frac{1}{16\pi^2(d-4)}\ \mbox{tr} \int d^4 x \sqrt{-g} <
Flat + WR +\qquad\qquad
\nonumber\\
\vphantom{\frac{1}{2}}
+ SR + SSR + FF + FR + RR>,
\end{eqnarray}

\noindent
where

\begin{eqnarray}\label{all}
&&
Flat =
\frac{1}{4} \hat S^4 - \hat W \hat S^2 + \frac{1}{2} \hat W^2
+ \hat S \hat N - \hat M
-\frac{1}{4} (L-1)(L-2) \nabla_\mu \hat S\ \nabla_\nu \hat S^{\mu\nu}
\nonumber\\
&&
+ \frac{1}{2} L(L-1) \nabla_\mu \hat S\ \nabla_\nu \hat S ^\nu \hat K^\mu
- \frac{1}{2} \nabla_\mu \hat S\ \nabla_\nu \hat S \Delta^{\mu\nu}
+ \frac{1}{3} \left(
\vphantom{\frac{1}{2}}
(L-1) \nabla_\mu \hat S ^\mu \hat S^2
\right.
\nonumber\\
&&
\vphantom{\frac{1}{2}}
- L \nabla_\mu \hat S\ \hat K^\mu \hat S^2
- (L-1) \nabla_\mu \hat S\ \hat S\ \hat S^\mu
\left.
+ L\ \nabla_\mu \hat S\ \hat S^2 \hat K^\mu
\right)
- L \nabla_\mu \hat S \hat W \hat K^\mu
\nonumber\\
&&
+ (L-2) \nabla_\mu \hat S \hat W ^\mu;\\
\nonumber\\
&&
WR =
-\ \frac{1}{2} L^2 \hat W\ \hat F_{\mu\nu} (Kn)^\mu \hat K ^\nu
+ \frac{1}{3} L\ \hat W\ \hat K^\alpha \Delta^{\mu\nu} n_\sigma
R^\sigma{}_{\mu\alpha\nu}
+ \frac{1}{3} L^2(L-1)
\nonumber\\
&&
\times \hat W\ \hat K ^{\mu\nu} \hat K^\alpha n_\sigma
R^\sigma{}_{\mu\alpha\nu}
- \frac{1}{6}(L-2)(L-3) \hat W^{\mu\nu} R_{\mu\nu};\\
\nonumber\\
&&
SR =
-\ \frac{1}{6} L^2(L-1) \hat S\ \nabla_\mu \hat F_{\alpha\nu} (Kn)^{\mu\nu}
\hat K^\alpha
+ \frac{2}{3} L \hat S\ \nabla_\mu \hat F_{\nu\alpha} (Kn)^\alpha
\Delta^{\mu\nu}
\nonumber\\
&&
- \frac{1}{12} (L-1)(L-2)(L-3)\hat S^{\alpha\mu\nu} \nabla_\alpha R_{\mu\nu}
- \frac{1}{12} L^2(L-1)(L-2) \hat S \hat K^{\mu\nu\alpha} \hat K^\beta
\nonumber\\
&&
\times n_\sigma \nabla_\alpha R^\sigma{}_{\mu\beta\nu}
+ L(L-1) \hat S\ \hat K^{\mu\nu} \Delta^{\alpha\beta} n_\sigma
\left(\frac{5}{12} \nabla_\alpha R^\sigma{}_{\nu\beta\mu} -
\frac{1}{12} \nabla_\mu R^\sigma{}_{\alpha\nu\beta}\right)
\nonumber\\
&&
- \frac{1}{2} L\ \hat S\ \hat K^\beta \Delta^{\mu\nu\alpha}
n_\sigma \nabla_\alpha R^\sigma{}_{\mu\beta\nu};\\
\nonumber\\
&&
SSR =
- \frac{1}{2} L(L-1) \hat S \hat S^\mu \hat F_{\mu\nu} \hat K^{\nu}
+ \frac{1}{2} L^2 \hat S^2 \hat F_{\mu\nu} (Kn)^{\mu} \hat K^\nu
+ \frac{1}{6} \hat S^2 \Delta^{\mu\nu}  R_{\mu\nu}
\nonumber\\
&&
+ \frac{1}{3} L(L-1) \hat S \hat S^\mu \hat K^\nu R_{\mu\nu}
+ \frac{1}{12} (L-1)(L-2) \hat S \hat S^{\mu\nu} R_{\mu\nu}
- \frac{1}{3} L^2 (L-1) \hat S^2
\nonumber\\
&&
\times \hat K^{\mu\nu} \hat K^\alpha
n_\sigma R^\sigma{}_{\mu\alpha\nu}
- \frac{1}{6} L(L-1)(L-2) \hat S \hat S^{\mu\nu}
\hat K^\alpha n_\sigma R^\sigma{}_{\mu\alpha\nu}
+ \frac{1}{3} (L-1) \hat S
\nonumber\\
&&
\times \hat S^\alpha \Delta^{\mu\nu} n_\sigma
R^\sigma{}_{\mu\alpha\nu}
- \frac{1}{3} L \hat S^2 \hat K^\alpha \Delta^{\mu\nu}
n_\sigma R^\sigma{}_{\mu\alpha\nu};\\
\nonumber\\
&&
FF = -\ \frac{1}{24} L^2 (L-1)^2 \hat K^{\mu\nu} F_{\mu\alpha}
\hat K^{\alpha\beta} F_{\nu\beta}
+ \frac{1}{24} L^2 \hat K^\mu F_{\beta\nu} \Delta^{\alpha\beta}
\hat K^\nu F_{\alpha\mu}
- \frac{5}{24} L^2
\nonumber\\
&&
\times \hat K^\mu F_{\beta\mu} \Delta^{\alpha\beta}
\hat K^\nu F_{\alpha\nu}
- \frac{1}{48} L^2 (L-1) \hat K^\mu F_{\beta\nu} \Delta^\nu
\hat K^{\alpha\beta} F_{\alpha\nu}
- \frac{1}{48} L^2 (L-1) \hat K^\mu
\nonumber\\
&&
\times
\vphantom{\frac{1}{2}}
F_{\beta\mu} \Delta^\nu \hat K^{\alpha\beta} F_{\alpha\nu};\\
\nonumber\\
&&
FR = \frac{1}{40} L^2 (L-1)(L-2) \Delta^\mu \hat K^\nu \hat
K^{\alpha\beta\gamma} F_{\mu\alpha} n_\sigma R^\sigma{}_{\gamma\beta\nu}
- L^2(L-1)(L-2)
\nonumber\\
&&
\times
\Delta^\nu \hat K^{\alpha\beta\gamma} \hat K^\mu n_\sigma
\left(
\frac{1}{60} R^\sigma{}_{\beta\gamma\mu} F_{\alpha\nu}
+ \frac{1}{12} R^\sigma{}_{\beta\gamma\nu} F_{\alpha\mu}
\right)
+ L^2 (L-1)^2 \Delta^\alpha \hat K^{\beta\gamma} \hat K^{\mu\nu}
\nonumber\\
&&
\times
n_\sigma
\left(
\frac{1}{60} R^\sigma{}_{\beta\mu\gamma} F_{\alpha\nu}
+\frac{1}{20} R^\sigma{}_{\alpha\mu\gamma} F_{\nu\beta}
+\frac{1}{15} R^\sigma{}_{\gamma\mu\alpha} F_{\nu\beta}
+\frac{1}{60} R^\sigma{}_{\mu\nu\gamma} F_{\alpha\beta}
\right)
+ L^2
\nonumber\\
&&
\times (L-1) \Delta^{\alpha\beta} \hat K^{\gamma\delta} \hat K^{\mu} n_\sigma
\left(
\frac{4}{15} R^\sigma{}_{\delta\beta\gamma} F_{\alpha\mu}
- \frac{1}{30} R^\sigma{}_{\beta\delta\alpha} F_{\gamma\mu}
- \frac{1}{15} R^\sigma{}_{\alpha\gamma\mu} F_{\beta\delta}
\right.
\nonumber\\
&&
\left.
- \frac{1}{30} R^\sigma{}_{\gamma\alpha\mu} F_{\beta\delta}
\right)
+ L^2 (L-1) \Delta^{\alpha\beta} \hat K^\gamma \hat K^{\mu\nu} n_\sigma
\left(
\frac{7}{60} R^\sigma{}_{\alpha\beta\mu} F_{\gamma\nu}
- \frac{11}{60} R^\sigma{}_{\beta\mu\gamma}
\right.
\nonumber\\
&&
\times
\left.
F_{\alpha\nu}
+ \frac{1}{5} R^\sigma{}_{\mu\alpha\gamma} F_{\beta\nu}
+ \frac{1}{60} R^\sigma{}_{\mu\alpha\nu} F_{\gamma\beta}
\right)
+ L^2 \Delta^{\mu\alpha\beta} \hat K^\gamma \hat K^\nu n_\sigma
\left(\frac{7}{20} R^\sigma{}_{\alpha\gamma\beta} F_{\nu\mu}
\right.
\nonumber\\
&&
\left.
+ \frac{1}{10} R^\sigma{}_{\alpha\beta\nu} F_{\gamma\mu}\right);\\
\nonumber\\
&&
\label{rr}
RR=
\frac{1}{10} L^2 \hat K^\delta \Delta^{\mu\nu\alpha\beta}
\hat K^\gamma n_\sigma n_\rho R^\sigma{}_{\alpha\beta\gamma}
R^\rho{}_{\mu\nu\delta}
+ L^2 (L-1)^2 (L-2) \hat K^{\beta\gamma\delta} \Delta^\alpha
\nonumber\\
&&
\times
\hat K^{\mu\nu} n_\sigma n_\rho
\left(
\frac{2}{45} R^\rho{}_{\alpha\delta\nu} R^\sigma{}_{\beta\mu\gamma}
-\frac{1}{120} R^\rho{}_{\delta\alpha\nu} R^\sigma{}_{\beta\mu\gamma}
\right)
+ L^2 (L-1) \hat K^\delta \Delta^{\alpha\beta\gamma}\hat K^{\mu\nu}
\nonumber\\
&&
\times
n_\sigma n_\rho
\left(
-\frac{1}{10} R^\rho{}_{\mu\gamma\nu} R^\sigma{}_{\alpha\delta\beta}
+\frac{1}{15} R^\rho{}_{\delta\alpha\nu} R^\sigma{}_{\beta\mu\gamma}
+\frac{1}{60} R^\rho{}_{\beta\delta\nu} R^\sigma{}_{\gamma\mu\alpha}
\right)+ L^2
\nonumber\\
&&
\times (L-1)^2 \hat K^{\gamma\delta} \Delta^{\alpha\beta}
\hat K^{\mu\nu} n_\sigma n_\rho
\left(
-\frac{1}{20} R^\rho{}_{\mu\beta\nu} R^\sigma{}_{\delta\alpha\gamma}
+\frac{1}{180} R^\rho{}_{\alpha\nu\beta} R^\sigma{}_{\gamma\delta\mu}
-\frac{7}{360}
\right.
\nonumber\\
&&
\left.
\times R^\rho{}_{\mu\gamma\nu} R^\sigma{}_{\alpha\delta\beta}
-\frac{1}{240} R^\rho{}_{\delta\beta\nu} R^\sigma{}_{\gamma\alpha\mu}
-\frac{1}{120} R^\rho{}_{\beta\gamma\nu} R^\sigma{}_{\alpha\delta\mu}
-\frac{1}{30} R^\rho{}_{\delta\beta\nu} R^\sigma{}_{\alpha\gamma\mu}
\right)
\nonumber\\
&&
\vphantom{\frac{1}{2}}
+ L^2 (L-1) (L-2)\ \hat K^\delta \Delta^{\mu\nu}
\hat K^{\alpha\beta\gamma} n_\sigma n_\rho
\left(
- \frac{1}{30} R^\rho{}_{\gamma\nu\beta} R^\sigma{}_{\alpha\delta\mu}
- \frac{1}{180} R^\rho{}_{\mu\gamma\nu}
\right.
\nonumber\\
&&
\left.
\times
R^\sigma{}_{\alpha\beta\delta}
+ \frac{1}{180} R^\rho{}_{\mu\gamma\delta} R^\sigma{}_{\alpha\beta\nu}
\right)
+ L^2 (L-1)^2 (L-2)\ \hat K^{\mu\nu} \Delta^{\delta}
\hat K^{\alpha\beta\gamma} n_\sigma n_\rho
\nonumber\\
&&
\times
\left(
\frac{1}{45} R^\rho{}_{\mu\gamma\nu} R^\sigma{}_{\alpha\beta\delta}
- \frac{1}{80} R^\rho{}_{\beta\nu\gamma} R^\sigma{}_{\mu\alpha\delta}
+ \frac{1}{90} R^\rho{}_{\beta\nu\gamma} R^\sigma{}_{\delta\alpha\mu}
\right)
+ L^2 (L-1) \hat K^{\mu\nu}
\nonumber\\
&&
\times
\Delta^{\alpha\beta\gamma}\hat K^\delta n_\sigma n_\rho
\left(
\frac{7}{120} R^\rho{}_{\beta\gamma\nu} R^\sigma{}_{\mu\alpha\delta}
- \frac{3}{40} R^\rho{}_{\beta\gamma\delta} R^\sigma{}_{\mu\alpha\nu}
+ \frac{1}{120} R^\rho{}_{\delta\gamma\nu} R^\sigma{}_{\alpha\beta\mu}
\right)
\nonumber\\
&&
+ L^2 (L-1)(L-2) \hat K^{\alpha\beta\gamma} \Delta^{\mu\nu}
\hat K^\delta n_\sigma n_\rho
\left(
- \frac{1}{24} R^\rho{}_{\mu\gamma\nu} R^\sigma{}_{\alpha\beta\delta}
- \frac{1}{180} R^\rho{}_{\nu\gamma\delta}
\right.
\nonumber\\
&&
\left.
\times R^\sigma{}_{\alpha\beta\mu}
- \frac{1}{360} R^\rho{}_{\delta\gamma\nu} R^\sigma{}_{\alpha\beta\mu}
\right)
- \frac{1}{120} L^2 (L-1)(L-2)(L-3) \hat K^{\mu\nu\alpha\beta}
\Delta^{\delta}
\nonumber\\
&&
\times
\hat K^\gamma n_\sigma n_\rho R^\rho{}_{\alpha\beta\gamma}
R^\sigma{}_{\mu\nu\delta}
- \frac{1}{80} L^2 (L-1)^2 (L-2)(L-3) \hat K^{\alpha\beta\gamma\delta}
\hat K^{\mu\nu} n_\sigma n_\rho
\nonumber\\
&&
\times
R^\rho{}_{\beta\gamma\mu} R^\sigma{}_{\alpha\delta\nu}
+ L^2 \hat K^\mu \Delta^{\alpha\beta\gamma}
\hat K^\nu n_\rho
\left(
- \frac{1}{8} R_{\beta\gamma} R^\rho{}_{\nu\alpha\mu}
+ \frac{3}{20} R_{\beta\gamma} R^\rho{}_{\mu\alpha\nu}
+ \frac{3}{40}
\right.
\nonumber\\
&&
\left.
\times R_{\alpha\mu} R^\rho{}_{\beta\gamma\nu}
+ \frac{1}{40} R^\sigma{}_{\beta\gamma\mu} R^\rho{}_{\nu\alpha\sigma}
- \frac{3}{20} R^\sigma{}_{\alpha\beta\mu} R^\rho{}_{\gamma\nu\sigma}
+ \frac{1}{10} R^\sigma{}_{\alpha\beta\nu} R^\rho{}_{\gamma\mu\sigma}
\right)
\nonumber\\
&&
+L^2 (L-1)\hat K^\gamma \Delta^{\alpha\beta} \hat K^{\mu\nu} n_\rho
\left(
\frac{1}{20} R_{\alpha\nu} R^\rho{}_{\gamma\beta\mu}
+ \frac{1}{20} R_{\alpha\gamma} R^\rho{}_{\mu\beta\nu}
+ \frac{1}{10} R_{\alpha\beta} R^\rho{}_{\mu\gamma\nu}
\right.
\nonumber\\
&&
\left.
+ \frac{1}{20} R^\sigma{}_{\alpha\nu\gamma} R^\rho{}_{\sigma\beta\mu}
- \frac{1}{60} R^\sigma{}_{\mu\alpha\nu} R^\rho{}_{\beta\sigma\gamma}
+ \frac{1}{10} R^\sigma{}_{\alpha\beta\gamma} R^\rho{}_{\mu\sigma\nu}
- \frac{1}{12} R^\sigma{}_{\alpha\beta\nu} R^\rho{}_{\mu\sigma\gamma}
\right)
\nonumber\\
&&
+ L^2 (L-1)^2 \hat K^{\alpha\beta} \Delta^{\gamma} \hat K^{\mu\nu} n_\rho
\left(
\frac{1}{60} R_{\alpha\mu} R^\rho{}_{\beta\nu\gamma}
- \frac{1}{20} R_{\alpha\mu} R^\rho{}_{\gamma\nu\beta}
\right.
+ \frac{1}{120} R_{\alpha\beta}
\nonumber\\
&&
\times
\left.
R^\rho{}_{\mu\nu\gamma}
+ \frac{3}{40} R_{\alpha\gamma} R^\rho{}_{\nu\beta\mu}
+ \frac{1}{20} R^\sigma{}_{\gamma\mu\alpha} R^\rho{}_{\nu\sigma\beta}
+ \frac{1}{120} R^\sigma{}_{\alpha\mu\gamma} R^\rho{}_{\beta\nu\sigma}
- \frac{1}{40} R^\sigma{}_{\alpha\mu\gamma}
\right.
\nonumber\\
&&
\left.
\times R^\rho{}_{\sigma\nu\beta}
+ \frac{1}{40} R^\sigma{}_{\alpha\mu\beta} R^\rho{}_{\sigma\nu\gamma}
- \frac{1}{20} R^\sigma{}_{\alpha\mu\beta} R^\rho{}_{\gamma\nu\sigma}
- \frac{1}{40} R^\sigma{}_{\mu\beta\nu} R^\rho{}_{\gamma\sigma\alpha}
\right)
+ L^2
\nonumber\\
&&
\vphantom{\frac{1}{2}}
\times
(L-1) \hat K^{\alpha\beta} \Delta^{\mu\nu} \hat K^{\gamma} n_\rho
\left(
\frac{1}{20} R^\sigma{}_{\mu\nu\beta} R^\rho{}_{\gamma\sigma\alpha}
- \frac{7}{60} R^\sigma{}_{\beta\mu\alpha} R^\rho{}_{\gamma\nu\sigma}
+ \frac{1}{20} R^\sigma{}_{\beta\mu\alpha}
\right.
\nonumber\\
&&
\left.
\times R^\rho{}_{\sigma\nu\gamma}
+ \frac{1}{10} R^\sigma{}_{\mu\beta\gamma} R^\rho{}_{\nu\alpha\sigma}
+ \frac{1}{60} R^\sigma{}_{\beta\mu\gamma} R^\rho{}_{\alpha\nu\sigma}
+ \frac{7}{120} R_{\alpha\beta} R^\rho{}_{\nu\gamma\mu}
+ \frac{11}{60} R_{\beta\mu}
\right.
\nonumber\\
&&
\left.
\vphantom{\frac{1}{2}}
\times R^\rho{}_{\nu\alpha\gamma}
\right)
+ L^2 (L-1) (L-2) \hat K^{\alpha\beta\gamma} \Delta^{\mu} \hat K^{\nu}
n_\rho
\left(
\frac{7}{240} R_{\alpha\beta} R^\rho{}_{\gamma\mu\nu}
+ \frac{7}{240} R_{\alpha\nu}
\right.
\nonumber\\
&&
\left.
\times R^\rho{}_{\beta\gamma\mu}
- \frac{1}{60} R_{\alpha\mu} R^\rho{}_{\beta\gamma\nu}
- \frac{1}{24} R^\sigma{}_{\alpha\beta\nu} R^\rho{}_{\sigma\gamma\mu}
+ \frac{1}{15} R^\sigma{}_{\alpha\beta\nu} R^\rho{}_{\mu\gamma\sigma}
+ \frac{1}{40} R^\sigma{}_{\alpha\beta\mu}
\right.
\nonumber\\
&&
\left.
\times R^\rho{}_{\sigma\gamma\nu}
+ \frac{1}{40} R_{\beta\gamma} R^\rho{}_{\nu\mu\alpha}
+ \frac{1}{48} R^\sigma{}_{\beta\gamma\mu} R^\rho{}_{\nu\alpha\sigma}
\right)
+ L^2 (L-1)^2 (L-2) \hat K^{\alpha\beta\gamma}
\nonumber\\
&&
\times \hat K^{\mu\nu} n_\rho
\left(
- \frac{7}{240} R_{\alpha\mu} R^\rho{}_{\beta\gamma\nu}
+ \frac{1}{240} R_{\beta\gamma} R^\rho{}_{\mu\alpha\nu}
- \frac{1}{40} R^\sigma{}_{\alpha\mu\beta} R^\rho{}_{\nu\gamma\sigma}
\right)
+ L (L-1)
\nonumber\\
&&
\times (L-2) (L-3) \hat K^{\mu\nu\alpha\beta}
\left(
\frac{1}{180} R_{\mu\nu} R_{\alpha\beta}
+ \frac{7}{720} R^\sigma{}_{\alpha\beta\rho} R^\rho{}_{\mu\nu\sigma}
\right),
\end{eqnarray}

\noindent
and an exact form of $\Delta^\mu, \Delta^{\mu\nu}, \Delta^{\mu\nu\alpha}$
and $\Delta^{\mu\nu\alpha\beta}$ are given in (\ref{Delta}).

We should note, that deriving (\ref{answer}) we omitted total derivatives,
and therefore we must make the following substitutions

\begin{eqnarray}
R_{\alpha\beta\mu\nu} R^{\alpha\beta\mu\nu}
= 2 R_{\alpha\beta\mu\nu} R^{\alpha\mu\beta\nu}
\rightarrow 4 R_{\mu\nu} R^{\mu\nu} - R^2.
\end{eqnarray}

Although the result is very large, the calculations of the effective action
can be easily made by computers. For example, we considered particular
cases (see next section) using tensor package \cite{tensor} for the REDUCE
analytical calculation system \cite{reduce}.


\section{Applications}
\hspace{\parindent}

The presented algorithm can be applied for calculations of the effective
action for theories regularized by higher derivatives \cite{slavnov} and
theories in the nonminimal gauges. May be there are some other prospects.
Here we consider only the simplest test examples in order to check the
correctness of the method. We prove, that our results agree with the earlier
known ones and each other.

\subsection{Arbitrary minimal operator}
\hspace{\parindent}

An arbitrary minimal operator of order $2l$ can be considered as a
particular case of a nonminimal one, if
$L = 2l$ and $K^{\mu\nu\ldots\alpha}{}_i{}^j =
K_0{}^{\mu\nu\ldots\alpha}\delta_i{}^j$, where $K_0$ was defined in
(\ref{k0}). Substituting it to (\ref{answer}) and taking into account
(\ref{angle}), we obtain (\ref{diaganswer2}).

So, we proved the agreement of the results for the minimal and nonminimal
operators.

\subsection{The minimal second order operator}
\hspace{\parindent}

If $l=1$ the operator (\ref{minimal1}) takes the form

\begin{equation}
D_{i}{}^{j} = \delta_{i}{}^{j} \Box + S^\mu{}_{i}{}^{j} \nabla_\mu +
W_{i}{}^{j}.
\end{equation}

Using the following concequences of (\ref{angle})

\begin{eqnarray}\label{middle}
&&<1> = 1,\qquad
<n_\mu n_\nu> = \frac{1}{4} g_{\mu\nu}, \\
&&<n_\mu n_\nu n_\alpha n_\beta> =
\frac{1}{24} (g_{\mu\nu} g_{\alpha\beta} +
g_{\mu\alpha} g_{\nu\beta} + g_{\mu\beta} g_{\nu\alpha}),\nonumber
\end{eqnarray}

\noindent
it is easy to see that (\ref{diaganswer1}) gives then the following
well-known result \cite{thooft}:

\begin{eqnarray}\label{secondorder}
&&\Gamma^{(1)}_\infty = \frac{1}{16\pi^2(d-4)}
 \mbox{tr} \int d^4 x \sqrt{-g}
\left(\frac{1}{12} Y_{\mu\nu} Y^{\mu\nu}
\right.
\nonumber\\
&&\hspace{5cm}
\left.
+ \frac{1}{2} X^2
+\frac{1}{60} R_{\mu\nu} R^{\mu\nu}
- \frac{1}{180} R^2 \right),
\end{eqnarray}

\noindent
where

\begin{eqnarray}
&&Y_{\mu\nu} = \frac{1}{2} \nabla_\mu S_\nu -
\frac{1}{2} \nabla_\mu S_\nu + \frac{1}{4} S_\mu S_\nu -
\frac{1}{4} S_\nu S_\mu + F_{\mu\nu},\nonumber\\
&&X = W - \frac{1}{2} \nabla_\mu S^\mu - \frac{1}{4} S_\mu S^\mu +
\frac{1}{6} R.
\end{eqnarray}


\subsection{The minimal forth order operator}
\hspace{\parindent}

A minimal forth order operator $(l=2)$ has the form

\begin{eqnarray}\label{forthoper1}
D_{i}{}^{j} = \delta_{i}{}^{j} \Box^2 +
S^{\mu\nu\alpha}{}_{i}{}^{j}\ \nabla_\mu \nabla_\nu \nabla_\alpha +
W^{\mu\nu}{}_{i}{}^{j}\ \nabla_\mu \nabla_\nu
+ N^{\mu}{}_{i}{}^{j}\ \nabla_{\mu} + M_{i}{}^{j}.
\end{eqnarray}

From (\ref{diaganswer1}) we obtained the result, that coincided with
the one found in \cite{barv} by Barvinsky and Vilkovisky up to the total
derivatives. (We do not presented it here because it is too large)

\subsection{Vector field}
\hspace{\parindent}

As another example we consider the vector field operator

\begin{equation}\label{vector}
D = \Box\ \delta_{\alpha}{}^{\beta} - \lambda\ \nabla_\alpha \nabla^\beta
+ P_{\alpha}{}^\beta,\qquad
\mbox{where}\qquad P_{\alpha\beta} = P_{\beta\alpha}.
\end{equation}

In order to rewrite it in the form (\ref{nonminimal}), we should
make the second term symmetric in $\alpha$ and $\beta$ by the commutation
of covariant derivatives. Then we found, that

\begin{eqnarray}
&&D_\alpha{}^\beta = \left(g^{\mu\nu} \delta_{\alpha}{}^{\beta} -
\frac{\lambda}{2}\ (g^{\mu\beta} \delta_\alpha{}^\nu +
g^{\nu\beta} \delta_\alpha{}^\mu)\right) \nabla_\mu \nabla_\nu +
P_{\alpha}{}^\beta
+ \frac{\lambda}{2}\ R_{\alpha}{}^\beta.
\end{eqnarray}

\noindent
So, nonzero coefficients are

\begin{eqnarray}
&&
K^{\mu\nu}{}_\alpha{}^\beta = g^{\mu\nu} \delta_{\alpha}{}^{\beta} -
\frac{\lambda}{2}\ (g^{\mu\beta} \delta_\alpha{}^\nu +
g^{\nu\beta} \delta_\alpha{}^\mu);\nonumber\\
&&
W_\alpha{}^\beta =
P_{\alpha}{}^\beta + \frac{\lambda}{2}\ R_{\alpha}{}^\beta
\end{eqnarray}

\noindent
and, therefore,

\begin{eqnarray}
&&
(Kn)_\alpha{}^\beta =
\delta_\alpha{}^\beta - \lambda\ n_\alpha n^\beta;\qquad
(Kn)^{-1}{}_\alpha{}^\beta = \delta_\alpha{}^\beta + \gamma\ n_\alpha
n^\beta,
\end{eqnarray}

\noindent
where ${\displaystyle \gamma = \frac{\lambda}{1-\lambda}}$.

After substituting it to (\ref{answer}) we found the following result

\begin{eqnarray}\label{vectoranswer}
&&
\Gamma^{(1)}_\infty=
\frac{1}{16\pi^2(d-4)} \int d^4 x\ \sqrt{-g}
\ \left(
\left(\frac{1}{24} \gamma^2 + \frac{1}{4} \gamma +\frac{1}{2}\right)
P_{\mu\nu} P^{\mu\nu}
+ \frac{1}{48} \gamma^2 P^2
\right.
\nonumber\\
&&
+ \left(\frac{1}{12} \gamma^2 +
\frac{1}{3} \gamma\right) R_{\mu\nu} P^{\mu\nu}
+ \left(\frac{1}{24} \gamma^2
+ \frac{1}{12} \gamma + \frac{1}{6}\right) R P
+ \left(\frac{1}{24} \gamma^2 + \frac{1}{12} \gamma
\right.
\nonumber\\
&&
\left.
-\frac{4}{15}\right) R_{\mu\nu} R^{\mu\nu}
\left.
+ \left(\frac{1}{48} \gamma^2 +
\frac{1}{12} \gamma +\frac{7}{60}\right) R^2\right),
\end{eqnarray}

\noindent
that is in agreement with \cite{barv} and \cite{fradkin}.
(Here $P \equiv P_\alpha{}^\alpha$.)

In order to check the result (\ref{answer}) we calculated one-loop
counterterms for the squared operator (\ref{vector}), that is a
nonminimal operator of the forth order:

\begin{eqnarray}
&&D^2{}_\alpha{}^\beta = \delta_\alpha{}^\beta \Box^2
- \lambda \nabla_\alpha \nabla^\beta \Box + 2 P_\alpha{}^\beta \Box
- \lambda \Box \nabla_\alpha \nabla^\beta
+ \lambda^2 \nabla_\alpha \Box \nabla^\beta
\vphantom{\frac{1}{2}}
\nonumber\\
&&
- \lambda P_\alpha{}^\mu \nabla_\mu \nabla^\beta
- \lambda P_\mu{}^\beta \nabla_\alpha \nabla^\mu
+ (\Box P_\alpha{}^\beta) + 2 (\nabla_\mu P_\alpha{}^\beta) \nabla^\mu
- \lambda (\nabla_\alpha \nabla^\mu P_\mu{}^\beta)
\vphantom{\frac{1}{2}}
\nonumber\\
&&
- \lambda (\nabla_\alpha P_\mu{}^\beta) \nabla^\mu
- \lambda (\nabla_\mu P^{\mu\beta}) \nabla^\alpha
+ P_\alpha{}^\mu P_\mu{}^\beta.
\vphantom{\frac{1}{2}}
\end{eqnarray}

By commuting covariant derivatives it can be rewritten in the form
(\ref{nonminimal}), where

\begin{eqnarray}
&&K^{\mu\nu\gamma\delta}{}_\alpha{}^\beta =
\delta_\alpha{}^\beta \frac{1}{3}\left(
g^{\mu\nu} g^{\gamma\delta}
+ g^{\mu\gamma} g^{\nu\delta}
+ g^{\mu\delta} g^{\nu\gamma}\right)
+ \frac{1}{12} (-2\lambda +\lambda^2)
\left(
g^{\mu\nu} \delta_\alpha{}^\gamma g^{\beta\delta}
\vphantom{\frac{1}{2}}
\right.\nonumber\\
&&
\vphantom{\frac{1}{2}}
+ g^{\mu\nu} \delta_\alpha{}^\delta g^{\beta\gamma}
+ g^{\mu\gamma} \delta_\alpha{}^\nu g^{\beta\delta}
+ g^{\mu\gamma} \delta_\alpha{}^\delta g^{\beta\nu}
+ g^{\mu\delta} \delta_\alpha{}^\nu g^{\beta\gamma}
+ g^{\mu\delta} \delta_\alpha{}^\gamma g^{\beta\nu}
+ g^{\nu\gamma} \delta_\alpha{}^\mu
\nonumber\\
&&
\left.
\times g^{\beta\delta}
+ g^{\nu\gamma} \delta_\alpha{}^\delta g^{\beta\mu}
+ g^{\nu\delta} \delta_\alpha{}^\mu g^{\beta\gamma}
+ g^{\nu\delta} \delta_\alpha{}^\gamma g^{\beta\mu}
+ g^{\gamma\delta} \delta_\alpha{}^\mu g^{\beta\nu}
+ g^{\gamma\delta} \delta_\alpha{}^\nu g^{\beta\mu}
\vphantom{\frac{1}{2}}
\right);\nonumber\\
\\
&&S^{\mu\nu\gamma}{}_\alpha{}^\beta = 0;\\
\nonumber\\
&&W^{\mu\nu}{}_\alpha{}^\beta=
2 P_\alpha{}^\beta g^{\mu\nu}
- \frac{\lambda}{2} P_\alpha{}^\mu g^{\nu\beta}
- \frac{\lambda}{2} P_\alpha{}^\nu g^{\mu\beta}
- \frac{\lambda}{2} P^{\beta\mu} \delta_\alpha{}^\nu
- \frac{\lambda}{2} P^{\beta\nu} \delta_\alpha{}^\mu
\nonumber\\
&&
- \frac{2}{3} R^{\mu\nu} \delta_\alpha{}^\beta;
+ \frac{1}{6} (\lambda - 2 \lambda^2)
\left(
  R_\alpha{}^\mu g^{\nu\beta}
+ R_\alpha{}^\nu g^{\mu\beta}
+ R^{\beta\mu} \delta_\alpha{}^\nu
+ R^{\beta\nu} \delta_\alpha{}^\mu
\right)
\nonumber\\
&&
+ \frac{1}{6} (2\lambda - \lambda^2)
\left(R_\alpha{}^{\mu\beta\nu}+R_\alpha{}^{\nu\beta\mu}\right)
+ \frac{1}{2} (2\lambda - \lambda^2)\ g^{\mu\nu} R_\alpha{}^\beta
\\
\nonumber\\
&&M_\alpha{}^\beta = P_{\alpha\mu} P^{\mu\beta}
+ \frac{\lambda}{2} P_{\alpha\mu} R^{\mu\beta}
+ \frac{\lambda}{2} P_{\mu\nu} R^\mu{}_\alpha{}^{\nu\beta}
+ \frac{1}{4} (2\lambda - \lambda^2)
R_{\alpha\mu\nu\gamma} R^{\gamma\mu\nu\beta}
\nonumber\\
&&
+ \frac{1}{12} (4 \lambda + 7 \lambda^2)\ R_{\mu\alpha\nu}{}^\beta
R^{\mu\nu}
+ \frac{1}{6} (\lambda - 2 \lambda^2)
R_{\alpha\mu} R^{\mu\beta}
- \frac{1}{2} R_{\mu\nu\gamma\alpha} R^{\mu\nu\gamma\beta}.
\end{eqnarray}

\noindent
It is easy to see that in this case
$(Kn)^{-1}{}_\alpha{}^\beta = \delta_\alpha{}^\beta
+ (2\gamma + \gamma^2)\ n_\alpha n^\beta$.
Substituting this expressions to (\ref{answer}) we obtain
by explicit calculation that an identity
$\mbox{tr}\ \ln\ D^2 = 2\ \mbox{tr}\ \ln\ D$ is satisfied.

\subsection{Gravity theory. The $\lambda$-family of gauge conditions}
\hspace{\parindent}

The gravitational field is described by the action

\begin{equation}
S = \int d^4x\ \sqrt{-g}\ R,
\end{equation}

\noindent
Its second variation can be found by making a substitution
$g_{\mu\nu} \rightarrow g_{\mu\nu} + h_{\mu\nu}$ and retaining terms
quadratic in $h_{\mu\nu}$. Due to the invariance under the general
coordinate transformations $x^\mu \rightarrow x^\mu + \xi^\mu$
$(Kn)_{\alpha\beta}{}^{\mu\nu}$ has a zero mode

\begin{eqnarray}
h_{\mu\nu} \rightarrow h_{\mu\nu} +
n_\mu \xi_\nu + n_\nu \xi_\mu,
\end{eqnarray}

\noindent
that should be deleted by adding to the action gauge fixing terms

\begin{equation}
S_{gf} = -\ \frac{1}{2} \int d^4x\ \sqrt{-g}\ g_{\mu\nu} \chi^\mu \chi^\nu,
\end{equation}

\noindent
where

\begin{equation}
\chi^\mu = \frac{1}{\sqrt{1+\lambda}} \left(g^{\mu\alpha} \nabla^\beta
h_{\alpha\beta} - \frac{1}{2} g^{\alpha\beta} \nabla^\mu h_{\alpha\beta}
\right).
\end{equation}

Then the second variation of the action takes the form

\begin{eqnarray}
&&\frac{\delta^2 S}{\delta h_{\alpha\beta}\ \delta h_{\mu\nu}} =
\sqrt{-g}\ C^{\alpha\beta,\gamma\delta} \left(
\vphantom{\frac{\lambda}{2(1+\lambda)}}
\delta_{\gamma\delta}^{\mu\nu} \Box
+ \frac{\lambda}{2(1+\lambda)} g^{\mu\nu}
\left(\nabla_\gamma\nabla_\delta+
\nabla_\delta\nabla_\gamma\right)
\right.
\nonumber\\
&&\left.
- \frac{\lambda}{2(1+\lambda)}
\left(\delta_\gamma^\mu \nabla_\delta \nabla^\nu
+ \delta_\gamma^\nu \nabla_\delta \nabla^\mu
+ \delta_\delta^\mu \nabla_\gamma \nabla^\nu
+ \delta_\delta^\nu \nabla_\gamma \nabla^\mu\right)
+ P_{\gamma\delta}{}^{\mu\nu}
\right),
\end{eqnarray}

\noindent
where

\begin{eqnarray}\label{p}
&&\delta^{\mu\nu}_{\alpha\beta} = \frac{1}{2} \left(
\delta_\alpha^\mu \delta_\beta^\nu +
\delta_\alpha^\nu \delta_\beta^\mu \right);\qquad
C^{\alpha\beta,\gamma\delta} = \frac{1}{4} \left(
g^{\alpha\gamma} g^{\beta\delta} + g^{\alpha\delta} g^{\beta\gamma}
- g^{\alpha\beta} g^{\gamma\delta} \right);\nonumber\\
&&P_{\gamma\delta}{}^{\mu\nu} =
R_\gamma{}^\mu{}_\delta{}^\nu + R_\gamma{}^\nu{}_\delta{}^\mu
+ \frac{1}{2} \left(\delta_\gamma^\mu R_\delta{}^\nu
+ \delta_\gamma^\nu R_\delta{}^\mu
+ \delta_\delta^\mu R_\gamma{}^\nu
+ \delta_\delta^\nu R_\gamma{}^\mu\right)
- g^{\mu\nu} R_{\gamma\delta}
\nonumber\\
&&
- g_{\gamma\delta} R^{\mu\nu}
- \delta^{\mu\nu}_{\gamma\delta} R
+ \frac{1}{2} g_{\gamma\delta} g^{\mu\nu} R.
\end{eqnarray}

$\mbox{tr}\ \ln\ (\sqrt{-g}\ C^{\alpha\beta,\mu\nu})$ gives zero contribution
in the dimensional regularization. So, in this case

\begin{eqnarray}
&&D_{\alpha\beta}{}^{\mu\nu} =
\vphantom{\frac{\lambda}{2(1+\lambda)}}
\delta_{\gamma\delta}^{\mu\nu} \Box - \frac{\lambda}{2(1+\lambda)}
\left(\delta_\gamma^\mu \nabla_\delta \nabla^\nu
+ \delta_\gamma^\nu \nabla_\delta \nabla^\mu
+ \delta_\delta^\mu \nabla_\gamma \nabla^\nu
+ \delta_\delta^\nu \nabla_\gamma \nabla^\mu\right)
\nonumber\\
&&
+ \frac{\lambda}{2(1+\lambda)} g^{\mu\nu}
\left(\nabla_\gamma\nabla_\delta+
\nabla_\delta\nabla_\gamma\right)
+ P_{\gamma\delta}{}^{\mu\nu}.
\end{eqnarray}

\noindent
and

\begin{eqnarray}
&&K^{\mu\nu}{}_{\alpha\beta}{}^{\gamma\delta} = g^{\mu\nu}
\delta_{\alpha\beta}{}^{\gamma\delta}
-\ \frac{\lambda}{4(1+\lambda)}
\left(
\delta_\alpha{}^\gamma \delta_\beta{}^\mu g^{\delta\nu}
+ \delta_\alpha{}^\gamma \delta_\beta{}^\nu g^{\delta\mu}
+ \delta_\alpha{}^\delta \delta_\beta{}^\mu g^{\gamma\nu}
\right.
\nonumber\\
&&
\vphantom{\frac{1}{2}}
+ \delta_\alpha{}^\delta \delta_\beta{}^\nu g^{\gamma\mu}
+ \delta_\beta{}^\gamma \delta_\alpha{}^\mu g^{\delta\nu}
+ \delta_\beta{}^\gamma \delta_\alpha{}^\nu g^{\delta\mu}
+ \delta_\beta{}^\delta \delta_\alpha{}^\mu g^{\gamma\nu}
\left.
+ \delta_\beta{}^\delta \delta_\alpha{}^\nu g^{\gamma\mu}
\right)
\nonumber\\
&&+ \frac{\lambda}{2(1+\lambda)} g^{\gamma\delta}
\left(\delta_\alpha{}^\mu \delta_\beta{}^\nu
+ \delta_\alpha{}^\nu \delta_\beta{}^\mu \right);\\
\nonumber\\
&&W_{\alpha\beta}{}^{\gamma\delta} =
P_{\alpha\beta}{}^{\gamma\delta}
-\ \frac{\lambda}{2(1+\lambda)}
\left(R_\alpha{}^\gamma{}_\beta{}^\delta
+ R_\alpha{}^\delta{}_\beta{}^\gamma\right)
+\ \frac{\lambda}{4(1+\lambda)}
\left(
\delta_\alpha{}^\gamma R_\beta{}^\delta
\right.
\nonumber\\
&&
\vphantom{\frac{1}{2}}
\left.
+ \delta_\alpha{}^\delta R_\beta{}^\gamma
+ \delta_\beta{}^\gamma R_\alpha{}^\delta
+ \delta_\beta{}^\delta R_\alpha{}^\gamma
\right);\\
\nonumber\\
&&(Kn)^{-1}{}_{\alpha\beta}{}^{\gamma\delta} =
\delta_{\alpha\beta}^{\gamma\delta}
+ \frac{\lambda}{2}
\left(
\delta_\alpha{}^\gamma n_\beta n^\delta
+ \delta_\alpha{}^\delta n_\beta n^\gamma
+ \delta_\beta{}^\gamma n_\alpha n^\delta
+ \delta_\beta{}^\delta n_\alpha n^\gamma
\right)
\nonumber\\
&&- \lambda g^{\gamma\delta} n_\alpha n_\beta.
\end{eqnarray}

The effective action is calculated by (\ref{answer}). For an arbitrary
$P_{\mu\nu}{}^{\alpha\beta}$, symmetric in each pair of indexes, we
found the following result:

\begin{eqnarray}
&&\frac{1}{16\pi^2(d-4)} \int d^4 x \sqrt{-g}
\ \left(
\frac{1}{48}
\lambda^2 (P^\mu{}_{\mu\nu}{}^\nu)^2
+(-4\lambda^2-12\lambda)\ P_{\mu\nu\alpha}{}^\alpha P_\beta{}^{\beta\mu\nu}
\right.
\nonumber\\
&&
\vphantom{\frac{1}{2}}
+(4\lambda^2+24\lambda+24) P_{\alpha\beta\mu\nu} P^{\mu\nu\alpha\beta}
+ 4\lambda^2 P_{\mu\nu\alpha\beta} P^{\nu\beta\mu\alpha}
+ 4 \lambda^2 P_{\mu\alpha\nu}{}^\alpha P_\beta{}^{\nu\beta\mu}
\nonumber\\
&&
\vphantom{\frac{1}{2}}
+ 2 \lambda^2 P^\alpha{}_{\alpha\mu\nu} P_\beta{}^{\beta\mu\nu}
- 8 \lambda^2 P_{\mu\alpha\nu}{}^\alpha P_\beta{}^{\beta\mu\nu}
+ (4 \lambda^2 + 8 \lambda + 8) P_{\mu\nu}{}^{\mu\nu} R
+ (- 4 \lambda^2
\nonumber\\
&&
\vphantom{\frac{1}{2}}
- 4 \lambda) P^\mu{}_{\mu\nu}{}^\nu R
+ 4\lambda^2 P^{\mu\nu\alpha}{}_\alpha R_{\mu\nu}
+ (8\lambda^2 + 32\lambda) P_\alpha{}^{\mu\alpha\nu} R_{\mu\nu}
+ (-4 \lambda^2 + 8\lambda)
\nonumber\\
&&
\times P_\alpha{}^{\alpha\mu\nu} R_{\mu\nu}
+ (-8\lambda^2 - 48\lambda) P_{\mu\nu\alpha\beta} R^{\mu\alpha\nu\beta}
+ (44\lambda^2 + 32\lambda
- 88)\ R_{\mu\nu} R^{\mu\nu}
\vphantom{\frac{1}{2}}
\nonumber\\
&&
\left.
\vphantom{\frac{1}{2}}
+ (-4\lambda^2+ 24\lambda + 28)\ R^2
\right).
\end{eqnarray}

Substituting here the expression (\ref{p}) for $P_{\mu\nu}{}^{\alpha\beta}$
we obtain

\begin{eqnarray}\label{maingr}
&&\frac{1}{16\pi^2(d-4)} \int d^4 x \sqrt{-g}
\left(\frac{1}{6} \left(4\lambda^2+4\lambda+7\right)
R^{\mu\nu} R_{\mu\nu}\right.
\nonumber\\
&&\hspace{7cm}
\left.
+ \frac{1}{12} \left(4\lambda^2+7\right) R^2 \right).
\end{eqnarray}

Moreover, we should add the contribution of Faddeev-Popov ghosts
with the Lagrangian density

\begin{equation}
L_{gh} = \bar c^\alpha \left(
\delta^\alpha{}_\beta \nabla^\mu \nabla_\mu + R{}^\alpha{}_\beta
\right) c^\beta.
\end{equation}

It can be easily found by (\ref{diaganswer2}). The result is

\begin{equation}\label{ghostgr}
-2\times \frac{1}{16\pi^2(d-4)} \int d^4 x\ \sqrt{-g}
\ \left(
\frac{7}{30} R_{\mu\nu} R^{\mu\nu} + \frac{17}{60} R^2 \right).
\end{equation}

The sum of (\ref{maingr}) and (\ref{ghostgr}) gives the ultimate
expression for the divergent part of the one-loop effective action

\begin{eqnarray}
&&
\Gamma^{(1)}_\infty=
\frac{1}{16\pi^2(d-4)} \int d^4 x\ \sqrt{-g}
\ \left(
\frac{1}{60} \left(20 \lambda^2+ 1\right) R^2
\right.
\nonumber\\
&&\hspace{5cm}
\left.
+ \frac{1}{30} \left(20\lambda^2+20\lambda+21\right)
R^{\mu\nu} R_{\mu\nu}
\right),
\end{eqnarray}

\noindent
that is in agreement with the results, found in \cite{kallosh} and
\cite{barv}.


\section{Conclusion}
\hspace{\parindent}

In this paper we calculate the divergent part of the one-loop effective
action for an arbitrary (minimal and nonminimal) operators without any
restrictions to their form and order in the curved space-time, using
t'Hooft-Veltman diagram technique \cite{thooft,1,2}.

Actually, we made some operations, that encounter in calculating Feynman
diagrams, namely, integration over a loop momentum, summation of all
divergent graphs and obtaining a manifestly covariant result by its weak
field limit in the general case. Then, in order to calculate the divergent
part of the effective action, one should only substitute the explicit
expression for second variation of an action.

Unfortunately, the master formula is very large and can hardly be used
for calculations without computers. Nevertheless, on the base of the
general algorithm we obtained one-loop counterterms for some examples,
namely, an arbitrary minimal operator, the vector field operator and
the gravity theory in the $\lambda$-gauge. Our results were in agreement
with the ones found earlier. The calculations were made by the tensor package for the REDUCE
analytical calculation system \cite{tensor}. For the considered examples
the required memory and execution time were the following (we used
IBM-486/DX-2/66/8Mb):

1. Calculation of the RR contribution for an arbitrary minimal operator
by (\ref{rr}) took 17 seconds. In this case the required memory was about
500 kb.

2. For the vector field operator and its square execution time was
174 and 515 seconds respectively. Required memory was 2 Mb.

3. The calculation of the one-loop counterterms for the gravity theory in
the $\lambda$-gauge took 170 minutes while the required memory was 8 Mb.

\vspace{1cm}

\noindent
{\Large\bf Acknolegments}

\vspace{1cm}

We are very grateful to professors D.V.Shirkov (JINR, Dubna) and
A.A.Slavnov (Steklov Mathematical Institute) on the attention to our
work, Dr. M.Yu.Kalmykov (JINR, Dubna) and Dr. V.V.Zhytnikov
(Moskow Pedagogical University, Moskow) on the valuable discussions.


\vspace{1cm}

\noindent
{\Large\bf Appendix}

\appendix

\section{Divergent diagrams calculation for a minimal operator}
\label{minimalcalc}
\hspace{\parindent}

Let us consider first logarithmically divergent graphs (1a)-(1e).
In order to find the divergent part of the diagram in this case
we should retain only terms without external momentums and perform
the remaining integration. As an example let us calculate a
diagram (1.d).

\begin{equation}
(1.d) = \frac{i}{2(2\pi)^4}
\mbox{tr} \int d^dk\ \frac{(Sk)\ (N(k-p))}{k^{2l} (k-p)^{2l}}.
\end{equation}

\noindent
Using the method described above, we can easily conclude that

\begin{eqnarray}\label{diag1}
(1.d)_\infty
= \frac{i}{2(2\pi)^4}\ \int d^dk\ \frac{1}{k^4} \mbox{tr} <\hat S\ \hat N>
= \frac{1}{16\pi^2(d-4)} \mbox{tr} <\hat S\ \hat N>,
\end{eqnarray}

\noindent
where

\begin{eqnarray}\label{notations}
&&\hat S = (Sn) = S^{\mu\nu\ldots \alpha} n_\mu n_\nu\ldots n_\alpha;
\nonumber\\
&&\vphantom{\frac{1}{2}}
\hat N = (Nn) = N^{\mu\nu\ldots \alpha} n_\mu n_\nu\ldots n_\alpha.
\end{eqnarray}

\noindent
and  $n_\mu = k_\mu/\sqrt{k^\alpha k_\alpha}$ is a unit vector. In a
similar fashion we have

\begin{eqnarray}\label{ln}
&&(1.a)_\infty = \frac{1}{64\pi^2(d-4)} \mbox{tr} <\hat S^4>;
\nonumber\\
&&(1.b)_\infty = -\ \frac{1}{16\pi^2(d-4)} \mbox{tr} <\hat W\ \hat S^2>;
\nonumber\\
&&(1.c)_\infty = \frac{1}{32\pi^2(d-4)} \mbox{tr} <\hat W^2>;
\nonumber\\
&&(1.e)_\infty = -\ \frac{1}{16\pi^2(d-4)} \mbox{tr} <\hat M>.
\end{eqnarray}

The calculation of linearly divergent graphs is a bit more difficult.
For example, in order to find the divergent part of the diagram

\begin{equation}
(1.f) = \frac{i}{2(2\pi)^4}
\mbox{tr} \int d^dk\ \frac{(Sk) (W(k-p))}{k^{2l} (k-p)^{2l}}
\end{equation}

\noindent
we should retain only terms linear in external momentum $p$. Using the
rule (\ref{eqa}) formulated in the appendix \ref{rules} we obtain

\begin{eqnarray}
&&\frac{1}{16\pi^2(d-4)}
\mbox{tr} <2l (pn) \hat S\ \hat W - (2l-2) p_\mu \hat W^\mu \hat S>
\nonumber\\
&&
= \frac{1}{32\pi^2(d-4)}
\mbox{tr}<(2l-1) p_\mu \hat S^\mu \hat W
- (2l-2)\ p_\mu \hat S\ \hat W^\mu>.
\end{eqnarray}

After a substitution $p_\mu \hat S \rightarrow - \partial_\mu \hat S$,
the result for this diagram takes the form

\begin{equation}
(1.f)_\infty = \frac{1}{32\pi^2(d-4)}
\mbox{tr} <- (2l-1) \partial_\mu \hat S^\mu \hat W +
(2l-2) \partial_\mu \hat S\ \hat W^\mu>.
\end{equation}

The second linearly divergent graph

\begin{equation}
(1.g) = -\ \frac{i}{6(2\pi)^4} \mbox{tr} \int d^dk
\ \frac{
\stackrel{\displaystyle (Sk)}{\scriptscriptstyle (-p)\phantom{+}}
\stackrel{\displaystyle (S(k+q))}{\scriptscriptstyle (-q)
\phantom{+qqqqqqq}}
\stackrel{\displaystyle (S(k-p))}{\scriptscriptstyle (p+q)
\phantom{+qqqqqqq}}}
{k^{2l}
(k-p)^{2l} (k+q)^{2l}}
\end{equation}

\noindent
can be calculated in the same way. (Here indexes in the bottom point
the argument of $S$). The result is

\begin{equation}
(1.g)_\infty = \frac{(2l-1)}{48\pi^2(d-4)}
\mbox{tr} <\partial_\mu \hat S^\mu \hat S\ \hat S -
\partial_\mu \hat S\ \hat S\ \hat S^\mu>.
\end{equation}

So, we should consider only the rest quadratically divergent diagram.

\begin{equation}
(1.h) = \frac{i}{4(2\pi)^4}
\mbox{tr} \int d^dk\ \frac{(Sk)(S(k-p))}{k^{2l}(k-p)^{2l}}.
\end{equation}

\noindent
Retaining logarithmically divergent terms we can easily find that

\begin{eqnarray}\label{diag8}
&&(1.h)_\infty =
\frac{1}{16\pi^2(d-4)} \mbox{tr} \int d^4 x
<- \frac{(2l-1)(l^2-1)}{2(2l+1)} \partial_\mu \hat S^{\mu\nu}
\partial_\nu \hat S
\nonumber\\
&&
+\frac{l^2}{2(2l+1)} \partial_\mu \hat S\ \partial^\mu \hat S
+\frac{(2l-1)^2 l}{4(2l+1)} \partial_\mu \hat S^\mu \partial_\nu
\hat S^\nu>.
\end{eqnarray}

The calculation of graphs in the curved space is much more difficult.
Here we consider as an example only the simplest group of diagrams,
namely, we find $a_3$ and $a_4$ coefficients in the equation (\ref{anzats})
by calculating graphs (2.a) and (2.b).

The vertex with $h_{\mu\nu}$ in (2a) should be found according to
(\ref{purt}) by series expansion of $\Box^l$ to the first order and
has the form

\begin{eqnarray}
\sum_{m=0}^{l-1} k^{2m} (k-p)^{2l-2m-2}
\left(-h^{\mu\nu} k_\mu (k-p)_\nu + \frac{1}{2} h^\alpha{}_\alpha
p^\mu (k-p)_\mu \right).
\end{eqnarray}

Then the graph (2.a) can be written as

\begin{eqnarray}\label{a3}
&&(2.a) = \frac{i}{2(2\pi)^4} \mbox{tr} \int d^d k\ \frac{1}{k^{2l} (k-p)^{2l}}
(Wk) \sum_{m=0}^{l-1} k^{2m} (k-p)^{2l-2m-2}
\nonumber\\
&&\times
\left(-h^{\mu\nu} k_\mu (k-p)_\nu + \frac{1}{2} h^\alpha{}_\alpha
p^\mu (k-p)_\mu \right).
\end{eqnarray}

It is easy to see that it is quadratically divergent. So, retaining terms
quadratic in external momentum $p$ we obtain the divergent part

\begin{eqnarray}
&&\frac{1}{16\pi^2 (d-4)} \mbox{tr} <
\hat W \left(
- \frac{l}{2} h^\alpha{}_\alpha p^2
- h^{\mu\nu} n_\mu n_\nu \left(
+ \frac{2}{3} l(l+1)(l+2) (n^\alpha p_\alpha)^2
\right.
\right.
\nonumber\\
&&
\left.
\left.
\vphantom{\frac{1}{2}}
- \frac{1}{2} l(l+1) p^2\right)
+l(l+1) (n^\gamma p_\gamma)\left(h^{\mu\nu} n_\mu p_\nu +
\frac{1}{2} h^\alpha{}_\alpha (n^\beta p_\beta)\right)
\right)>.
\end{eqnarray}

\noindent
Using rules formulated in the appendix \ref{rules}, it can be
written as

\begin{eqnarray}\label{w1}
&&(2.a)_\infty = \frac{1}{16\pi^2(d-4)}
\mbox{tr} < - \frac{1}{12} (2l-3)(2l-4)(2l-5) \hat W^{\mu\nu\alpha\beta}
h_{\alpha\beta} p_\mu p_\nu
\nonumber\\
&&
- \frac{1}{12} (l-1)(2l-3) \hat W^{\mu\nu}
(- p^2 h_{\mu\nu} - p_\mu p_\nu h^\alpha{}_\alpha
+ 2 p_\mu p_\alpha h^\alpha{}_\nu)
+ \frac{l}{6}\hat W(-h^\alpha{}_\alpha
\nonumber\\
&&\times
\vphantom{\frac{1}{2}}
p^2+ h^{\mu\nu} p_\mu p_\nu)>.
\end{eqnarray}

Nevertheless, (\ref{w1}) can not be presented as a weak field limit
of a covariant expression. The matter is that the graphs (2.a) and
(2.b) can not be considered separately.

The vertex in the tadpole diagram (2.b) can be found by series expansion of
$W^{\mu\nu\ldots\alpha}{}_i{}^j \nabla_\mu \nabla_\nu \ldots \nabla_\alpha$
in powers of $h_{\mu\nu}$ to the first order. Then, retaining only
nontrivial contributions, we have

\begin{eqnarray}\label{b3}
&&(2.b) = \frac{i}{2(2\pi)^4}\ \mbox{tr} \int d^d k
\sum_{m=0}^{2l-4} (2l-3-m) (k+p)_{\mu_1} \ldots (k+p)_{\mu_m}
\Gamma^{\alpha\ {\scriptscriptstyle (1)}}_{\mu_{m+1}\mu_{m+2}}
\nonumber\\
&&\times
k{}_\alpha k{}_{\mu_{m+3}} \ldots k_{\mu_{2l-2}}
\frac{1}{k^{2l}}
W{}^{\mu_1\mu_2 \ldots \mu_{2l-2}},
\end{eqnarray}

\noindent
where

\begin{equation}
\Gamma^{\alpha\ {\scriptscriptstyle (1)}}_{\beta\gamma} =
\frac{1}{2} (p_\beta h_\gamma{}^\alpha + p_\gamma h_\beta{}^\alpha
- p^\alpha h_{\beta\gamma})
\end{equation}

\noindent
is the Cristoffel symbol in the weak field limit.

By (\ref{rule2}) after simple transformations the divergent part
of (2.b) can be written as

\begin{eqnarray}\label{w2}
(2.b)_\infty = \frac{(2l-3)(2l-4)(2l-5)}{192\pi^2(d-4)}
\mbox{tr}<\hat W^{\mu\nu\alpha\beta} h_{\alpha\beta} p_\mu p_\nu>.
\end{eqnarray}

\noindent
The sum of (\ref{w1}) and (\ref{w2}) unlike each of the graphs
considered separately can be presented as a weak field approximation
of a covariant expression

\begin{eqnarray}
\frac{1}{16\pi^2(d-4)} \mbox{tr} \int d^4 x \sqrt{-g}
<\frac{l}{6}\hat W R - \frac{1}{6}(l-1)(2l-3) R_{\mu\nu} \hat W^{\mu\nu}>,
\end{eqnarray}

\noindent
that is the ultimate answer for the graphs (2.a) and (2.b).


\section{Divergent graphs calculation for nonminimal operator}
\label{nondiagcalc}
\hspace{\parindent}

The flat space divergent graphs are the same as for a minimal operator.
Nevertheless, the calculations are a bit different. As earlier, we begin
with the consideration of logarithmically divergent graphs. For example,

\begin{eqnarray}
&&(1.d)_\infty =
\frac{i}{2(2\pi)^4}\ \mbox{tr} \int d^dk\ (Sk)\ (Kk)^{-1} (N(k-p))
\left.
\vphantom{\frac{1}{2}}
(K(k-p))^{-1}
\right|_\infty
\nonumber\\
&&
= \frac{1}{16\pi^2(d-4)}\ \mbox{tr} <\hat S\ \hat N>.
\end{eqnarray}

\noindent
where

\begin{eqnarray}
&&\hat S \equiv (Kn)^{-1} (Sn);\qquad \hat N \equiv (Kn)^{-1} (Nn),
\qquad\mbox{an so on}
\end{eqnarray}

\noindent
(compare with (\ref{notations})!)

The other logarithmically divergent graphs can be calculated in the
same way. A form of the result coincides with (\ref{ln}). (But the
notations differ!)

Now let us consider linearly divergent graphs.

\begin{eqnarray}
(1.f) =\frac{i}{2(2\pi)^4} \mbox{tr} \int d^dk (Sk)(Kk)^{-1}
(W(k-p))(K(k-p))^{-1}.
\end{eqnarray}

We should expand this expression into series over external momentum
$p$ and retain logarithmically divergent terms (It is easy to see
that they are linear in $p$). The expansion of $(K(k-p))^{-1}$ is
found in the appendix \ref{rules}. By eq. (\ref{expan}) and (\ref{Delta})
we obtain, that

\begin{eqnarray}
(1.f)_\infty = \frac{1}{16\pi^2(d-4)} \mbox{tr} <
- L \partial_\mu \hat S\ \hat W
\ \hat K^\mu + (L-2) \partial_\mu \hat S\ \hat W^\mu>.
\end{eqnarray}

In a similar fashion we find the divergent parts of the diagrams
(1.g) and (1.h). The results are presented in (\ref{nondiagflat}).

As for a minimal operator we do not consider here all curved space
diagrams. In order to illustrate the method we present only the
calculation of the $WR$ contribution (or the graphs (2.a), (2.b),
(2.o), (2.p) and (3.a)). The other graphs are computed in the same
way, but the calculations are much more cumbersome.

If $K^{\mu\nu\ldots\alpha}$ does not depend on metric, the vertexes
with $h_{\mu\nu}$ should be obtained by the series expansion of
covariant derivatives in the operator (\ref{nonminimal}) to the first
order. Then, retaining only logarithmically divergent terms and
performing the integration as described in the section \ref{nonmin}, we
find the divergent part of a diagram.

By this method we obtain the following answer for the quadratically
divergent graph (2.a)

\begin{eqnarray}\label{a2}
&&(2.a)_\infty =
\frac{1}{16\pi^2(d-4)} \mbox{tr} < \frac{1}{3} L(L-1)(L-2) \hat W
\hat K^{\mu\nu\alpha}
p_\alpha \Gamma^\sigma_{\mu\nu} n_\sigma
\nonumber\\
&&
+ \frac{1}{2} L(L-1) \hat W \hat K^{\mu\nu}
\Gamma^\sigma_{\mu\nu} p_\sigma
+ \frac{1}{2} L (L-1) \hat W \hat K^{\mu\nu}
\Delta^\alpha \Gamma^\sigma_{\mu\nu} n_\sigma p_\alpha>,
\end{eqnarray}

\noindent
where $\Delta^\mu$ is the first coefficient in the propagator expansion.
Its explicit form is found in the appendix \ref{rules}.

Similarly we obtain

\begin{equation}
(2.b)_\infty = \frac{1}{6} (L-2)(L-3)(L-4)
< \hat W ^{\mu\alpha\beta} p_\mu \Gamma_{\alpha\beta}^\sigma n_\sigma>.
\end{equation}

\noindent
Using rules, derived in the appendix, it can be rewritten as

\begin{eqnarray}\label{b2}
&&
\frac{1}{16\pi^2(d-4)} \mbox{tr} <
\frac{1}{3}\left(p_\alpha \Gamma_{\mu\nu}^\sigma
+ p_\mu \Gamma_{\nu\alpha}^\sigma
+ p_\nu \Gamma_{\mu\alpha}^\sigma\right)
\hat W n_\sigma
\left(
\frac{1}{6}L(L-1)(L-2)
\right.
\nonumber\\
&&
\times
\hat K^{\mu\nu\alpha}
+ \frac{1}{2} L (L-1) \hat K^{\mu\nu} \Delta^\alpha
\left.
\vphantom{\frac{1}{2}}
+ L \hat K^\alpha \Delta^{\mu\nu}\right)
+ \frac{1}{3} \left(p_\sigma \Gamma_{\mu\nu}^\sigma
+ 2 p_\mu \Gamma_{\nu\sigma}^\sigma\right) \hat W
\nonumber\\
&&
\times \left(\frac{1}{2} L(L-1)\ \hat K_{\mu\nu}
+ L\ \hat K^\mu \Delta^\nu
\right)>.
\end{eqnarray}

Summing up the results (\ref{a2}) and (\ref{b2}), we obtain

\begin{eqnarray}\label{metric}
&&
\frac{1}{16\pi^2(d-4)}
\mbox{tr} <\frac{1}{2} L(L-1)(L-2) \hat W
\hat K^{\mu\nu\alpha} p_\alpha \Gamma^\sigma_{\mu\nu} n_\sigma
+\frac{1}{3} L (L-1) \hat W
\nonumber\\
&&
\times \hat K^{\mu\nu} \Delta^\alpha n_\sigma
\left(
2 p_\alpha \Gamma^\sigma_{\mu\nu} + p_\mu \Gamma^\sigma_{\nu\alpha}
\right)
+\frac{1}{3} L \hat W\ \hat K^\alpha \Delta^{\mu\nu} n_\sigma
\left(p_\alpha \Gamma^\sigma_{\mu\nu} + 2 p_\mu \Gamma^\sigma_{\nu\alpha}
\right)
\nonumber\\
&&
+ \frac{1}{3} L(L-1) \hat W\ \hat K^{\mu\nu}
\left(2 p_\sigma \Gamma^\sigma_{\mu\nu} + p_\mu \Gamma^\sigma_{\nu\sigma}
\right)
+ \frac{1}{3} L\ \hat W\ \hat K^\mu \Delta^\nu \left(p_\sigma
\Gamma^\sigma_{\mu\nu}
+ 2 p_\mu
\right.
\nonumber\\
&&
\vphantom{\frac{1}{2}}
\left.
\times \Gamma^\sigma_{\nu\sigma}\right)>.
\end{eqnarray}

Diagrams, containing the connection $\omega_{\alpha i}{}^j$ are
calculated in a similar fashion. (We remind, that $\omega_{\alpha i}^j$
should be considered as a weak field.) We should consider 2 linear
divergent graphs (2.o) and (2.p). By the same arguments as above, it is
easy to see, that

\begin{eqnarray}
&&
(2.o)_\infty = \frac{1}{16\pi^2(d-4)}
\mbox{tr} <- \frac{1}{2} L(L-1) \hat W \hat K^{\mu\nu}
\omega_\mu p_\nu - L \hat W \hat K^\mu \omega_\mu \Delta^\nu p_\nu >;
\nonumber\\
&&
(2.p)_\infty =-\ \frac{1}{16\pi^2(d-4)}\ \mbox{tr}
<L(L-1) \hat W^{\mu\nu} \omega_\mu p_\nu>
\nonumber\\
&&\hspace{5cm}
=-\ \frac{1}{16\pi^2(d-4)} \mbox{tr}
<\hat W\ p_\nu \omega_\mu \Delta^{\mu\nu}>.
\end{eqnarray}

So, the whole contribution of this diagrams is

\begin{eqnarray}\label{connection}
\frac{1}{16\pi^2(d-4)} \mbox{tr} <
- \frac{1}{2} L(L-1) \hat W \hat K^{\mu\nu} p_\nu \omega_\mu
- L\ \hat W \hat K^\mu p_\nu \omega_\mu \Delta^\nu\qquad\qquad
\nonumber\\
- \hat W p_\nu \omega_\mu \Delta^{\mu\nu}>.
\end{eqnarray}

As we mention above, (\ref{metric}) and (\ref{connection}) can not
be presented as a weak field limit of covariant expressions.
The matter is that we should take into account a contribution of
fields $\phi^b$. For this purpose we consider a quadratically divergent
diagram (3.a). It is written as

\begin{equation}
(3.a) = \frac{1}{16\pi^2(d-4)}
\mbox{tr} \int d^d k (Wk) (Kk)^{-1} \frac{\partial (K(k-p))}{\partial \phi^b}
\phi^b (K(k-p))^{-1}.
\end{equation}

Expanding the integrant into series over external momentum $p$
and retaining only terms, quadratic in it (they are logarithmically
divergent), we obtain

\begin{eqnarray}
&&(3.a)_\infty = \frac{1}{16\pi^2(d-4)}
\mbox{tr} < \frac{1}{2} L(L-1)
\hat W\ (Kn)^{-1} \frac{\partial  (Kn)^{\mu\nu}}
{\partial \phi^b} \phi^b p_\mu p_\nu
+ L\ \hat W
\nonumber\\
&&\vphantom{\frac{1}{2}}
\times (Kn)^{-1}
\frac{\partial (Kn)^\mu}{\partial \phi^b} \phi^b \Delta^\nu  p_\mu p_\nu
+ \hat W\ (Kn)^{-1}
\frac{\partial (Kn) }{\partial \phi^b} \phi^b \Delta^{\mu\nu} p_\mu p_\nu>,
\end{eqnarray}

\noindent
that can be written as

\begin{eqnarray}\label{phi}
&&\frac{1}{16\pi^2(d-4)}
\mbox{tr} < \frac{1}{2} L(L-1)
\hat W\ (Kn)^{-1} \partial_\nu \partial_\mu (Kn)^{\mu\nu}
+ L\ \hat W\ (Kn)^{-1}
\nonumber\\
&&
\vphantom{\frac{1}{2}}
\times \partial_\nu \partial_\mu (Kn)^\mu  \Delta^\nu
+ \hat W\ (Kn)^{-1} \partial_\nu \partial_\mu (Kn) \Delta^{\mu\nu}>.
\end{eqnarray}

Taking into account (\ref{covder}) and using rules, formulated in the
appendix \ref{rules}, we find additional terms

\begin{eqnarray}\label{covphi}
&&
(3.a)_\infty = -\ \frac{1}{16\pi^2(d-4)}
\mbox{tr} < \frac{1}{2} L(L-1) p_\nu \hat W \left(
(L-2) \Gamma_{\mu\alpha}^\sigma \hat K^{\mu\nu\alpha} n_\sigma
+ \Gamma_{\mu\alpha}^\mu
\right.
\nonumber\\
&&
\vphantom{\frac{1}{2}}
\left.
\times \hat K^{\nu\alpha}
+ \Gamma_{\mu\alpha}^\nu \hat K^{\mu\alpha}
+ (Kn)^{-1} \omega_{\mu} (Kn)^{\mu\nu} -
\hat K^{\mu\nu} \omega_\mu
\right)
+ L p_\nu \hat W \left((L-1)\Gamma_{\mu\alpha}^\sigma
\right.
\nonumber\\
&&
\vphantom{\frac{1}{2}}
\left.
\times \hat K ^{\mu\alpha} n_\sigma
+ \Gamma_{\mu\alpha}^\mu \hat K^\alpha +
(Kn)^{-1} \omega_\mu (Kn)^\mu
- \hat K^\mu \omega_\mu
\right) \Delta^\nu
+ p_\nu \hat W \left(L \Gamma_{\mu\alpha}^\sigma
\hat K^\alpha n_\sigma
\right.
\nonumber\\
&&
\vphantom{\frac{1}{2}}
\left.
+ (Kn)^{-1} \omega_\mu (Kn) - \hat K\ \omega_\mu \right) \Delta^{\mu\nu}>
=\nonumber\\
\nonumber\\
&&
= \frac{1}{16\pi^2(d-4)}
\mbox{tr} < \frac{1}{2} L(L-1) p_\nu \hat W \left(
(L-2) \Gamma_{\mu\alpha}^\sigma \hat K^{\mu\nu\alpha} n_\sigma
+ \Gamma_{\mu\alpha}^\mu \hat K^{\nu\alpha}
\right.
\nonumber\\
&&
\vphantom{\frac{1}{2}}
\left.
+ \Gamma_{\mu\alpha}^\nu \hat K^{\mu\alpha}
- \hat K^{\mu\nu} \omega_\mu
\right)
+ L p_\nu \hat W
\left((L-1)\ \Gamma_{\mu\alpha}^\sigma \hat K ^{\mu\alpha} n_\sigma
+ \Gamma_{\mu\alpha}^\mu \hat K^\alpha
- \hat K^\mu \omega_\mu
\right)
\nonumber\\
&&
\vphantom{\frac{1}{2}}
\times \Delta^\nu
+ p_\nu \hat W
\left(L \Gamma_{\mu\alpha}^\sigma \hat K^\alpha n_\sigma
- \omega_\mu \right) \Delta^{\mu\nu}>.
\end{eqnarray}

Adding (\ref{covphi}) to (\ref{metric}) and (\ref{connection}),
we obtain the following result in the weak field limit

\begin{eqnarray}
&&
\frac{1}{16\pi^2(d-4)} \mbox{tr}<
\frac{1}{3} L\ \hat W \hat K^\alpha \Delta^{\mu\nu} n_\sigma
\left(p_\alpha \Gamma^\sigma_{\mu\nu} - p_\nu \Gamma^\sigma_{\mu\alpha}
\right)
+ \frac{1}{6} L (L-1) \hat W \hat K^{\mu\nu}
\nonumber\\
&&
\times \Delta^\alpha n_\sigma
\left(p_\alpha \Gamma^\sigma_{\mu\nu} - p_\nu \Gamma^\sigma_{\mu\alpha}
\right)
+ \frac{1}{6} L(L-1) \hat W \hat K^{\mu\nu}
\left(p_\sigma\Gamma_{\mu\nu}^\sigma
- p_\mu \Gamma_{\nu\sigma}^\sigma\right)
- \frac{1}{6} L\ \hat W
\nonumber\\
&&
\times \hat K^\mu \Delta^\nu
\left(p_\sigma\Gamma_{\mu\nu}^\sigma
- p_\mu \Gamma_{\nu\sigma}^\sigma\right)
- \frac{1}{2} L^2 \hat W \left(p_\mu \omega_\nu - p_\nu \omega_\mu
\right)
\hat K^\mu \hat K^\nu>,
\end{eqnarray}

\noindent
that can be (unlike (\ref{metric}) and (\ref{connection}) !)
presented as a first term in the expansion of

\begin{eqnarray}\label{result}
&&
\frac{1}{16\pi^2(d-4)} \mbox{tr}<
-\ \frac{1}{2} L^2 \hat W F_{\mu\nu} \hat K^\mu \hat K^\nu
+ \frac{1}{6} L(L-1) \hat W \hat K^{\mu\nu} R_{\mu\nu}
+ \frac{1}{6} L^2 \hat W
\nonumber\\
&&
\times \hat K^\mu \hat K^\nu R_{\mu\nu}
+ \frac{1}{3} L\ \hat W \hat K^\alpha \Delta^{\mu\nu} n_\sigma
R^\sigma{}_{\mu\alpha\nu}
+ \frac{1}{6} L (L-1) \hat W \hat K^{\mu\nu} \Delta^\alpha n_\sigma
R^\sigma{}_{\mu\alpha\nu}>\nonumber\\
\end{eqnarray}

\noindent
in powers of $h_{\mu\nu}$ and $\omega_\mu{}_i{}^j$.

This expression is a final result for the considered group of diagrams.
Now we try to present it in the most compact form. Using rules, formulated
in the appendix, we find, that

\begin{eqnarray}
-\ \frac{1}{2} L(L-1) \hat W\ \hat K^{\mu\nu} R_{\mu\nu} +
L^2 \hat W\ \hat K^\mu \hat K^\nu R_{\mu\nu}\qquad\qquad\qquad
\nonumber\\
= \hat W \Delta^{\mu\nu} R_{\mu\nu} =
\frac{1}{2} (L-2)(L-3) \hat W^{\mu\nu} R_{\mu\nu}
\end{eqnarray}

Then, taking into account the following simple identities

\begin{eqnarray}
&&
\vphantom{\frac{1}{2}}
(\nabla_\mu \nabla_\nu - \nabla_\nu \nabla_\mu) (Kn) =
L n_\rho R^\rho{}_{\alpha\mu\nu} (Kn)^\alpha
+ F_{\mu\nu} (Kn) - (Kn) F_{\mu\nu};
\nonumber\\
&&
\vphantom{\frac{1}{2}}
(\nabla_\mu \nabla_\nu - \nabla_\nu \nabla_\mu) (Kn)^\beta =
(L-1) n_\rho R^\rho{}_{\alpha\mu\nu} (Kn)^{\alpha\beta} +
R^\beta{}_{\alpha\mu\nu} (Kn)^\alpha
\nonumber\\
&&
\vphantom{\frac{1}{2}}
+ F_{\mu\nu} (Kn)^\beta
- (Kn)^\beta F_{\mu\nu};
\nonumber\\
&&
\vphantom{\frac{1}{2}}
(\nabla_\mu \nabla_\nu - \nabla_\nu \nabla_\mu) (Kn)^{\beta\gamma} =
(L-2)\ n_\rho R^\rho{}_{\alpha\mu\nu} (Kn)^{\alpha\beta\gamma}
+ R^\beta{}_{\alpha\mu\nu} (Kn)^{\alpha\gamma}
\nonumber\\
&&
\vphantom{\frac{1}{2}}
+ R^\gamma{}_{\alpha\mu\nu} (Kn)^{\alpha\beta}
+ F_{\mu\nu} (Kn)^{\beta\gamma} - (Kn)^{\beta\gamma} F_{\mu\nu},
\end{eqnarray}

\noindent
we obtain the final result

\begin{eqnarray}
&&\frac{1}{16\pi^2(d-4)} \mbox{tr}<
-\ \frac{1}{2} L^2 \hat W\ \hat F_{\mu\nu} (Kn)^\mu \hat K ^\nu
+ \frac{1}{3} L\ \hat W\ \hat K^\alpha \Delta^{\mu\nu} n_\sigma
R^\sigma{}_{\mu\alpha\nu}
\nonumber\\
&&
+ \frac{1}{3} L^2(L-1) \hat W\ \hat K ^{\mu\nu} \hat K^\alpha n_\sigma
R^\sigma{}_{\mu\alpha\nu}
- \frac{1}{6}(L-2)(L-3) \hat W^{\mu\nu} R_{\mu\nu}>.
\end{eqnarray}

\noindent
The other diagrams are considered in the same way.


\section{Integration over angles}
\label{rules}
\hspace{\parindent}

Let us first us start with (\ref{angle}):

\begin{eqnarray}\label{angle1}
&&<n_{\mu_{1}} n_{\mu_{2}} \ldots n_{\mu_{2m}}>\ \equiv
\frac{1}{2^{m} (m+1)!}
\nonumber\\
\vphantom{\frac{1}{2}}
&&\times
\left(g_{\mu_{1}\mu_{2}} g_{\mu_{3}\mu_{4}} \ldots g_{\mu_{2m-1}\mu_{2m}}
+ \mbox{permutations of}\ (\mu_1 \ldots \mu_{2m})
\right)
\end{eqnarray}

\noindent
It can be interpreted in the following way:
In order to obtain the result of the angle integration we should
make pairs of $n_\alpha$ by all possible ways and add a numerical constant.
Each pair  of $n_\alpha$ and $n_\beta$ should be substituted by
$g_{\alpha\beta}$.

The sum contains $(2m-1)!!$ terms. Hence, if we contract (\ref{angle1})
with a totally symmetric tensor
$A^{\mu_1\mu_2\ldots\mu_{2m}} \equiv A_{(2m)}$
(here the bottom index points the tensor rank) the result will be

\begin{equation}\label{rule1}
<(A_{(2m)}n)> =  \frac{(2m-1)!!}{2^m(m+1)!}
A^{\mu_1\ldots\mu_m}{}_{\mu_1\ldots\mu_m},
\end{equation}

Similarly one can find that for a symmetric tensor $A_{(2m-1)}$ with
$2m-1$ indexes the following equation takes place:

\begin{eqnarray}\label{rule2}
&&<n_\alpha (A_{(2m-1)}n)> =  \frac{(2m-1)!!}{2^m(m+1)!}
A^{\mu_1\ldots\mu_{m-1}}{}_{\mu_1\ldots\mu_{m-1}\alpha}
\nonumber\\
&&\hspace{6cm}
= \frac{2m-1}{2(m+1)} <(A_{(2m-1)}n)_\alpha>,\hspace{1cm}
\end{eqnarray}

In the more general case we will use the following consequence of
(\ref{angle}):

\begin{eqnarray}
&&<n_{\mu_1} n_{\mu_2} \ldots n_{\mu_{2m}}> = \frac{1}{2(m+1)}
\left(\vphantom{\sqrt{1}}g_{\mu_1\mu_2}
<n_{\mu_3} n_{\mu_4} \ldots n_{\mu_{2m}}> \right. \\
&&+ \left. g_{\mu_1\mu_3}
<n_{\mu_2} n_{\mu_4} \ldots n_{\mu_{2m}}> + \ldots + g_{\mu_1\mu_{2m}}
<n_{\mu_2} n_{\mu_3} \ldots n_{\mu_{2m-1}}>\right).\nonumber
\end{eqnarray}

Making contraction with 2 symmetric tensors we find that

\begin{eqnarray}\label{eqa}
&&<n_\alpha (A_{(2m)}n) (B_{(2p-1)}n)> =
\frac{1}{2(m+p+1)}<\left(2m (A_{(2m)}n){}_\alpha (B_{(2p-1)}n)
\right.
\nonumber\\
&&
\left.
+ (2p-1) (A_{(2m)}n) (B_{(2p-1)}n){}_\alpha\right)>.
\vphantom{\frac{1}{2}}
\end{eqnarray}

\noindent
This equation can be easily generalized to a greater number of
symmetric tensors.

The rules (\ref{rule1}), (\ref{rule2}) and (\ref{eqa}) are used in
calculating Feynman graphs for a minimal operator. For a nonminimal
operator we should formulate different identities.

Let us consider an integral

\begin{equation}
\int d^d k\ f(k)
\end{equation}

\noindent
and assume, that all terms in it are more than logarithmically divergent.
Therefore in the dimensional regularization it is equal to 0.

A substitution $k_\mu \rightarrow k_\mu+p_\mu$ do not change the
divergent part of the integral. Nevertheless, on the other hand,
we can calculate it explicitly, using the method described above.
For example, let us consider that $f(k)$ transforms as

1. $f(k) \rightarrow \alpha^{-3} f(k)$ if $k_\mu \rightarrow \alpha k_\mu$.
Then, retaining only logarithmically divergent terms, we obtain

\begin{equation}\label{dil1}
0 =\left(\int\ d^dk\ f(k+p)\right)_\infty =
-\ \frac{2i\pi^2}{d-4} <\delta_\mu f(n)> p^\mu,
\end{equation}

\noindent
if

\begin{equation}
f(k+p) \equiv f(k) + \delta_\mu f(k) p^\mu + \delta_{\mu\nu} f(k)
p^\mu p^\nu + \ldots.
\end{equation}

\noindent
($\delta_{\mu\nu\ldots}$ must be considered as symmetric in its indexes)

From (\ref{dil1}) we have

\begin{equation}
<\delta_\mu f(n)> = 0.
\end{equation}

2. If $f(k)\ g(k) \rightarrow \alpha^{-3} f(k)\ g(k)$ we find

\begin{equation}
\left(\int\ d^dk\ g(k)\ f(k+p)\right)_\infty =
\left(\int\ d^dk\ g(k-p)\ f(k)\right)_\infty
\end{equation}

\noindent
and therefore

\begin{equation}\label{rule3}
<g(n)\ \delta_\mu f(n)>\ = -\ <\delta_\mu g(n)\ f(n)>.
\end{equation}

3. If $f(k) g(k) \rightarrow \alpha^{-2} f(k) g(k)$, using an identity

\begin{eqnarray}
\left(\int\ d^dk\ g(k) f(k+p)\right)_\infty =
\left(\int\ d^dk\ g(k-p) f(k)\right)_\infty\nonumber
\end{eqnarray}

\noindent
it is easy to see, that

\begin{equation}
<g(n) \delta_{\mu\nu} f(n)> = <\delta_{\mu\nu} g(n) f(n)>.
\end{equation}

\noindent
Other cases are considered in a similar fashion.

In order to apply this rules for calculating Feynman graphs, we should
know coefficients of the propagator expansion in powers of the external
momentum

\begin{eqnarray}\label{expan}
&&
\vphantom{\frac{1}{k^4}}
(K(k+p))^{-1} =
(Kk)^{-1} + \delta^\mu (Kk)^{-1} p_\mu +
\delta^{\mu\nu} (Kk)^{-1} p_\mu p_\nu
+ \delta^{\mu\nu\alpha} (Kk)^{-1}
\nonumber\\
&&
\times p_\mu p_\nu p_\alpha + \ldots
\equiv
\left(1 + \frac{1}{k} \Delta^\mu p_\mu +
\frac{1}{k^2} \Delta^{\mu\nu} p_\mu p_\nu
+ \frac{1}{k^3} \Delta^{\mu\nu\alpha}
p_\mu p_\nu p_\alpha
+ \frac{1}{k^4}
\right.
\nonumber\\
&&
\left.
\vphantom{\frac{1}{k^4}}
\times \Delta^{\mu\nu\alpha\beta} p_\mu p_\nu p_\alpha
p_\beta
+ \ldots \right) (Kk)^{-1}.
\end{eqnarray}

The coefficient $\Delta^\mu$ -- $\Delta^{\mu\nu\alpha\beta}$ can be
found by expanding the identity

\begin{equation}
1_i{}^j = (K(k+p))_i{}^m (K(k+p))^{-1}{}_m{}^j
\end{equation}

\noindent
in powers of $p_\mu$. The result is

\begin{eqnarray}\label{Delta}
&&
\vphantom{\frac{1}{2}}
\Delta^\mu = - L \hat K^\mu;
\nonumber\\
&&
\Delta^{\mu\nu} \equiv -\ \frac{1}{2} L(L-1) \hat K^{\mu\nu} +
L^2 \hat K^{(\mu} \hat K^{\nu)};
\nonumber\\
&&
\Delta^{\mu\nu\alpha} =
-\ \frac{1}{6} L(L-1)(L-2) \hat K^{\mu\nu\alpha}
+ \frac{1}{2} L^2(L-1) \hat K^{(\mu\nu} \hat K^{\alpha)}
+ \frac{1}{2} L^2 (L-1)
\nonumber\\
&&
\qquad
\vphantom{\frac{1}{2}}
\times \hat K^{(\alpha)} \hat K^{\mu\nu)}
- L^3 \hat K^{(\mu} \hat K^\nu \hat K^{\alpha)};
\nonumber\\
&&
\Delta^{\mu\nu\alpha\beta} =
-\ \frac{1}{24} L(L-1)(L-2)(L-3) \hat K^{\mu\nu\alpha\beta}
+ \frac{1}{6} L^2 (L-1) (L-2) \hat K^{(\mu\nu\alpha}
\nonumber\\
&&
\qquad
\times
\hat K^{\beta)}
+ \frac{1}{6} L^2 (L-1)(L-2) \hat K^{(\beta} \hat K^{\mu\nu\alpha)}
+ \frac{1}{4} L^2 (L-1)^2 \hat K^{(\mu\nu} \hat K^{\alpha\beta)}
\nonumber\\
&&
\qquad
- \frac{1}{2} L^3 (L-1) \hat K^{(\mu\nu} \hat K^\alpha \hat K^{\beta)}
- \frac{1}{2} L^3 (L-1) \hat K^{(\alpha} \hat K^{\mu\nu} \hat K^{\beta)}
- \frac{1}{2} L^3 (L-1)
\nonumber\\
&&
\qquad
\times \hat K^{(\alpha} \hat K^\beta \hat K^{\mu\nu)}
+ L^4 \hat K^{(\mu} \hat K^\nu \hat K^\alpha \hat K^{\beta)},
\vphantom{\frac{1}{2}}
\end{eqnarray}

\noindent
where

\begin{equation}
A^{(i_1 i_2 \ldots i_n)} \equiv \frac{1}{n!} \left(A^{i_1 i_2 \ldots i_n}
+ A^{i_2 i_1 \ldots i_n} + \ldots + A^{i_n i_{n-1} \ldots i_1} \right).
\end{equation}

\noindent
(Other coefficients is not needed for the calculation of the divergent
part of the effective action).

The coefficient $\Delta^\mu$ -- $\Delta^{\mu\nu\alpha\beta}$ satisfy
the following useful identities

\begin{eqnarray}
&&
\Delta^{\mu\nu} + \frac{1}{2} L(L-1) \hat K^{\mu\nu} + L \hat K^{(\mu}
\Delta^{\nu)} = 0;
\nonumber\\
&&
\Delta^{\mu\nu\alpha} + \frac{1}{6} L(L-1)(L-2) \hat K^{\mu\nu\alpha}
+ \frac{1}{2} L(L-1) \hat K^{(\mu\nu}\Delta^{\alpha)}
+ L \hat K^{(\mu}\Delta^{\nu\alpha)} = 0;
\nonumber\\
&&
\Delta^{\mu\nu\alpha\beta}
+ \frac{1}{24} L(L-1)(L-2)(L-3) \hat K^{\mu\nu\alpha\beta}
+ \frac{1}{6} L(L-1)(L-2) \hat K^{(\mu\nu\alpha} \Delta^{\beta)}
\nonumber\\
&&
+ \frac{1}{2} L(L-1) \hat K^{(\mu\nu}\Delta^{\alpha\beta)}
+ L \hat K^{(\mu}\Delta^{\nu\alpha\beta)} = 0
\end{eqnarray}

\noindent
and so on.

Now we illustrate formulated rules by the simplest example. Let us
consider

\begin{equation}
<(L-3) (Nn)^\alpha (Kn)^{-1}>,
\end{equation}

\noindent
where $N^{\mu\nu\ldots\alpha}$ is a totally symmetric tensor with $L-3$
indexes. $(L-3) (Nn)^\alpha$ can be presented as $\delta^\alpha (Nn)$.
Using (\ref{rule3}), we obtain

\begin{equation}
<(L-3) (Nn)^\alpha (Kn)^{-1}> = - <(Nn)\ \delta^\alpha (Kn)^{-1}>.
\end{equation}

\noindent
Substituting here $\Delta^\alpha$ we find an identity

\begin{equation}
<(L-3) \hat N^\alpha> =  L <\hat N \hat K^\alpha >.
\end{equation}

\noindent
Other cases can be considered in the same way.


\noindent
\hspace*{-3cm}

\noindent
\hspace*{-3cm}

\noindent
\hspace*{-3cm}

\end{document}